\pdfoutput=1
\documentclass[11pt,a4paper,twoside]{article}

\usepackage[showcomments]{ajtex}
\DeclareRobustCommand{\DIEP}{\ensuremath{%
    \mathchoice{\includegraphics[height=2ex]{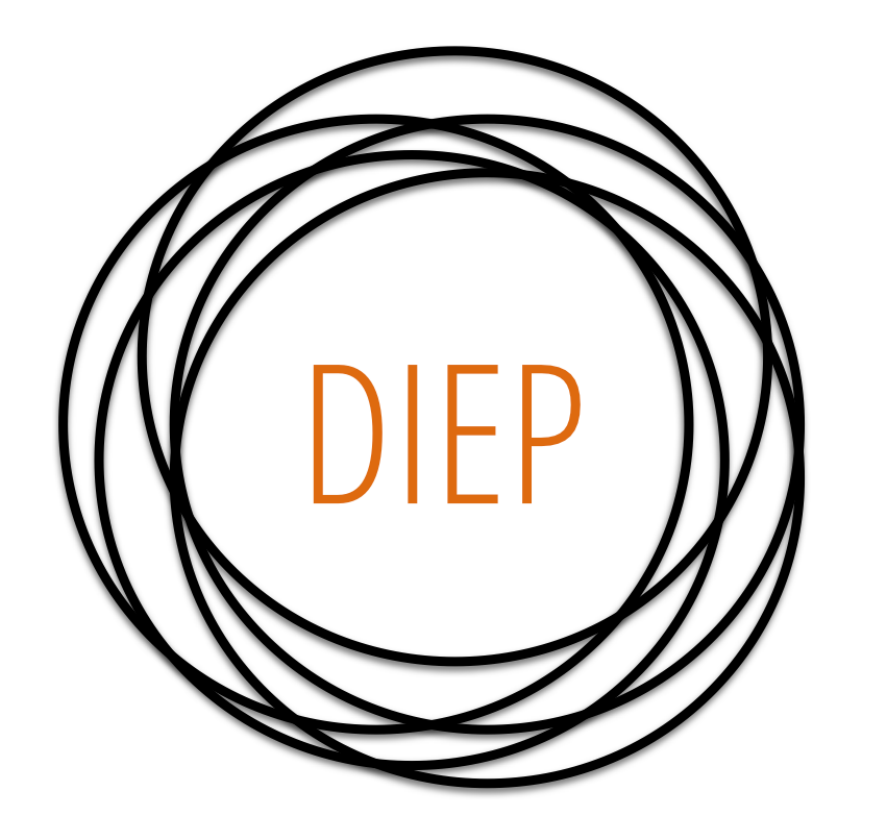}}
    {\includegraphics[height=2ex]{DIEPs.pdf}}
    {\includegraphics[height=1.5ex]{DIEPs.pdf}}
    {\includegraphics[height=1ex]{DIEPs.pdf}}
  }}

\newcommand{\bs}{\boldsymbol}

\title{A stable and causal model of magnetohydrodynamics}

\author[a,\DIEP]{Jay Armas}\email{j.armas@uva.nl}

\author[b,c]{Filippo Camilloni}\email{filippo.camilloni@nbi.ku.dk}

\affiliation[a]{Institute for Theoretical Physics, University of Amsterdam, 1090
  GL Amsterdam, The Netherlands}

\affiliation[\DIEP]{Dutch Institute for Emergent Phenomena, 1090 GL Amsterdam, The Netherlands}

\affiliation[b]{Dipartimento di Fisica e Geologia, Universit\`a di Perugia, I.N.F.N. Sezione di Perugia, \\ Via Pascoli, I-06123 Perugia, Italy}

\affiliation[c]{Niels Bohr Institute, Copenhagen University,\\ Blegdamsvej 17, DK-2100 Copenhagen \O{}, Denmark} 

\abstract{We formulate the theory of first-order dissipative magnetohydrodynamics in an arbitrary hydrodynamic frame under the assumption of parity-invariance and discrete charge symmetry. We study the mode spectrum of Alfvén and magnetosonic waves as well as the spectrum of gapped excitations and derive constraints on the transport coefficients such that generic equilibrium states with constant magnetic fields are stable and causal under linearised perturbations. We solve these constraints for a specific equation of state and show that there exists a large family of hydrodynamic frames that renders the linear fluctuations stable and causal. This theory does not require introducing new dynamical degrees of freedom and therefore is a promising and simpler alternative to M\"{u}ller-Israel-Stewart-type theories. Together with a detailed analysis of transport, entropy production and Kubo formulae, the theory presented here is well suited for studying dissipative effects in various contexts ranging from heavy-ion collisions to astrophysics.}

\begin{document}

\maketitle



\section{Introduction}
Magnetohydrodynamics (MHD) is the theory describing the long-wavelength dynamics of dynamical electromagnetic fields coupled to matter at finite temperature. The broad interest in MHD stems from the fact that it is the standard theory describing aspects of plasma physics \cite{anile_1990}, heavy-ion collisions \cite{Gursoy:2014aka}, the large scale structure of the universe \cite{Brandenburg:1996fc, refId0} and the dynamics of classes of black holes in particular regimes \cite{Grozdanov:2017kyl}, to mention a few. In the specific context of astrophysics, modelling of relativistic plasmas is useful for understanding neutron star mergers \cite{Abbott_2017}, relativistic jets and flaring activity \cite{10.1093/mnras/stab163, Ripperda:2021zpn}, besides black hole accretion \cite{Abramowicz:2011xu, 10.1093/mnras/stw166}. Any novel insight into the structure and foundations of MHD is bound to have a wide range of applications. 

Hydrodynamics is an effective theory that can be constructed order-by-order in a gradient expansion, constrained by symmetry and the second law of thermodynamics. MHD is not an exception. In the past years, MHD has been formulated using these same principles under the usual assumption that the fluid/plasma is electrically neutral at hydrodynamic length scales and electric fields are Debye screened \cite{Schubring:2014iwa, Grozdanov:2016tdf, Hernandez:2017mch, Armas:2018ibg, Armas:2018atq, Armas:2018zbe}. It has been shown, in particular, that MHD can be recast in terms of conservation laws for a stress tensor and a two-form current in such a way that it can be viewed as a theory of one-form superfluidity \cite{Armas:2018atq, Armas:2018zbe}. This point of view provides a simpler formulation of MHD as it avoids the unnecessary work of dealing with electric fields, which are algebraically determined in terms of the remaining hydrodynamic and magnetic fields \cite{Armas:2018atq, Armas:2018zbe}. 

However, issues that are encountered in usual formulations of hydrodynamics are directly inherited in analogous formulations of MHD. The most severe of these issues, is the fact that conventional relativistic hydrodynamics, à la  Landau and Lifshitz \cite{landau1987fluid} or Eckart \cite{PhysRev.58.919}, predicts an unstable equilibrium state and acausal signal propagation \cite{Hiscock:1985zz, Hiscock:1987zz}. In turn, the most popular resolution of these issues is known as the M\"{u}ller-Israel-Stewart (MIS) theory \cite{Muller:1967zza, Israel:1976tn, Israel:1979wp, ISRAEL1976213} which requires the addition of second order terms in gradients and the introduction of new degrees of freedom. Despite MIS being the standard theory used in modelling many systems of interest for the past 60 years, there are still open problems related to causality (e.g. at the nonlinear level) and several unwanted features (for instance, the inexistence of solutions describing shocks at high Mach number) - see e.g. \cite{Biswas:2020rps, Disconzi:2020ijk, Pandya:2021ief} for a recent discussion of these issues.  

Recently, it has been proposed that the problems of stability and superluminal propagation in relativistic hydrodynamics can be cured by a suitable choice of hydrodynamic variables out-of-equilibrium - known as a choice of frame. In particular, \cite{Bemfica:2017wps, Bemfica:2019knx, Kovtun:2019, Bemfica:2020zjp} (BDNK) identified a family of frames for which first order dissipative relativistic hydrodynamics is both stable and causal. This was subsequently applied to hydrodynamics with a conserved $U(1)$ current \cite{Hoult:2020eho} and to chiral hydrodynamics \cite{Speranza:2021bxf}.\footnote{In the non-relativistic case, choices of frames were studied in \cite{Poovuttikul:2019ckt, Armas:2020mpr}.} The authors of \cite{Pandya:2021ief} performed a numerical study of the first order formulation of BDNK for uncharged relativistic hydrodynamics and shown that it performs better or equally good as MIS for large classes of initial data in the case of conformal fluids. One of the major advantages of this approach is that there is no need to deal with second order hydrodynamics nor to introduce new degrees of freedom.

The purpose of this paper is to provide a family of frames that renders first order dissipative MHD causal and stable. This is not a trivial task due to the sheer number of transport coefficients involved and the different excitations/modes present in the system. Parity-violating MHD, without the presence of a conserved particle/baryon number, is characterised by 19 transport coefficients in the Landau frame (once microscopic CPT symmetry is imposed) \cite{Armas:2018zbe}. Restricting to parity-invariant MHD, 12 transport coefficients remain, while imposing discrete charge and parity symmetry leads to 7 independent transport coefficients at first order \cite{Armas:2018zbe}. It is this latter restricted case that we study in this paper. However, while keeping track of 7 coefficients may appear simple, when studying MHD in a general frame in this setting, we are required to work with 28 transport coefficients. In addition, we need to require stability and causality for three types of mode excitations (Alfvén waves, fast and slow magnetosonic waves) due to inherent anisotropy. Despite the increased difficulty of MHD compared to earlier similar studies of relativistic hydrodynamics, we are able to provide a method by which the desired family of frames can be obtained for a given equation of state. We expect this result to significantly contribute to the understanding of heavy-ion collisions, accretion disc dynamics and plasma physics.

This paper is organised as follows. In Sec.~\ref{sec:MHDframe} we formulate MHD in a general frame following \cite{Armas:2018atq, Armas:2018zbe}. In this section, we introduce several transport coefficients and obtain inequality constrains due to positivity of entropy production and microscopic CPT symmetry. In Sec.~\ref{sec:constraints} we study the general constraints imposed by causality and stability on parameter space and also show that the Landau frame is acausal and unstable. In Sec.~\ref{sec:EOS} we study in detail the constraints for a specific plasma. In Sec.~\ref{sec:discussion} we summarise our results and discuss future work. We also provide several appendices. App.~\ref{app:details1form} gives details about the formulation of MHD following \cite{Armas:2018atq, Armas:2018zbe}. App. ~\ref{app:Landauframe} contains the required formulae to translate our results into the Landau frame. App.~\ref{app:boostedframe} explains how to obtain dispersion relations at finite spatial velocity. In App.~\ref{app:furtherdetails} we give various details about the general constraints obtained in Sec.~\ref{sec:constraints}. Finally, in App.~\ref{app:discrete} we provide the transformations of various MHD fields under discrete symmetries. 

\section{Magnetohydrodynamics in a general frame}\label{sec:MHDframe}
In this section we review parity-invariant MHD as formulated in \cite{Armas:2018atq, Armas:2018zbe} and discuss first order dissipative corrections in the Landau frame. We then generalise the formulation of  \cite{Armas:2018atq, Armas:2018zbe} to arbitrary hydrodynamic frames imposing parity-invariance and discrete charge symmetry. We also perform a detailed analysis of frame transformations, entropy constraints and Kubo formulae.

\subsection{Equations of motion}
MHD describes the coupling of Maxwell's equations to thermal degrees of freedom of matter. We can describe this coupling in terms of the conservation of a stress tensor $T^{\mu\nu}$ that includes both fluid and electromagnetic fields together with Maxwell equations in the presence of an external current $J^\mu_{\text{ext}}$ such that
\begin{equation} \label{eq:eomtrad}
\nabla_\mu T^{\mu\nu}=F^{\mu\rho}J_\rho~~,~~J^\mu+J^\mu_{\text{ext}}=0~~,~~\nabla_{[\mu}F_{\nu\lambda]}=0~~,
\end{equation}
where the electromagnetic current $J^\mu$ is decomposed into a Maxwell part and a matter part such that $J^\mu=\nabla_\nu F^{\mu\nu}+J^\mu_{\text{matter}}$ where $J^\mu_{\text{matter}}$ is the current associated with matter fields and $F_{\mu\nu}$ is the electromagnetic field strength which we can decompose as
\begin{equation}
F_{\mu\nu}=2u_{[\mu}E_{\nu]}-\epsilon_{\mu\nu\rho\sigma}u^\rho B^\sigma~~,
\end{equation}
where $E^\mu$ and $B^\mu$ are the electric and magnetic fields respectively, $u^\mu$ is the four velocity and $\epsilon_{\mu\nu\rho\sigma}$ is the Levi-Civita tensor. Note that in \eqref{eq:eomtrad} we have not considered the existence of additional conserved particle/baryon number currents and we also did not consider radiation terms commonly discussed in accretion disc theory \cite{Abramowicz:2011xu}.

The system of equations \eqref{eq:eomtrad} is specified in terms of hydrodynamic fields via the dependence of $T^{\mu\nu}, J^\mu$ on the temperature $T$, electric chemical potential $\nu$ conjugate to electric charge density $q$ and fluid velocity $u^\mu$, normalised such that $u^\mu u_\mu=-1$ as well as $E^\mu$ and $B^\mu$. In particular, the stress tensor conservation law in \eqref{eq:eomtrad} provides dynamics for $T$ and $u^\mu$, Maxwell's equations $J^\mu+J^\mu_{\text{ext}}=0$ provides dynamics for $\nu$ and $E^\mu$ while the Bianchi identity $\nabla_{[\mu}F_{\nu\lambda]}=0$ gives dynamics to $B^\mu$. Treating MHD as an effective theory, we ascribe a derivative counting scheme to the various quantities. Specifically, $T\sim u^\mu\sim B^\mu\sim\mathcal{O}(1)$. However, the usual assumptions of MHD is that the plasma is electrically neutral at hydrodynamic length scales and that electric fields are Debye screened. In terms of the hydrodynamic expansion this amounts to take $J^\mu_{\text{ext}}\sim\mathcal{O}(\partial)$ which, generically, is equivalent to $\nu\sim\mathcal{O}(1)$ and $q\sim E^\mu\sim\mathcal{O}(\partial)$.\footnote{Taking $J^\mu_{\text{ext}}\sim\mathcal{O}(\partial)$ also implies that, generically, the electric charge density $q(T,\nu)$ appearing in the electric current in the usual formulation of MHD has the derivative ordering $q(T,\nu)\sim\mathcal{O}(\partial)$. As will be explained in greater detail in App.~\ref{app:details1form}, if discrete one-form charge and parity symmetries are imposed, we end up with configurations that satisfy $\nu\sim\mathcal{O}(\partial)$ and $q(T,\nu)\sim\mathcal{O}(\partial^2)$. This means that the fluid configurations we consider are electrically neutral up to order $\mathcal{O}(\partial)$. } As such, in order to solve the system \eqref{eq:eomtrad} it is only required to look at the first and last equations. Maxwell's equations are only needed in order to obtain expressions for $q$ and $E^\mu$ in terms of the remaining hydrodynamic fields. 

Given that Maxwell's equations are not necessary for the dynamics, it is possible to recast the remaining equations as conservation laws. To this end, we introduce a two-form anti-symmetric current $J^{\mu\nu}$ and a background three-form field strength $H_{\mu\nu\lambda}=3\partial_{[\mu}b_{\nu\lambda]}$ where $b_{\mu\nu}$ is a two-form gauge field such that
\begin{equation} \label{eq:Jdual}
J^{\mu\nu}=\frac{1}{2}\epsilon^{\mu\nu\lambda\rho}F_{\lambda\rho}~~,~~J^\mu_{\text{ext}}=\frac{1}{6}\epsilon^{\mu\nu\rho\sigma}H_{\nu\rho\sigma}~~.
\end{equation}
These identifications allow to recast the first and third equations in \eqref{eq:eomtrad} as
\begin{equation} \label{eq:eomdual}
\nabla_\mu T^{\mu\nu}=\frac{1}{2}H^{\nu\rho\sigma}J_{\rho\sigma}~~,~~\nabla_\mu J^{\mu\nu}=0~~.    
\end{equation}
Within this language, the dynamical degrees of freedom are $T, u^\mu$ and $B^\mu$ but it is convenient to split the degrees of freedom contained in $B^\mu$ in terms of a unit normalised vector $h^\mu=B^\mu/|B|+\mathcal{O}(\partial)$ satisfying $h^\mu h_\mu=1$, $h^\mu u_\mu=0$ where $|B|$ is the modulus of the magnetic field and a chemical potential $\mu=-2|B|\partial P/\partial |B|^2+\mathcal{O}(\partial)$ that accounts for the strength of the magnetic field and where $P$ is the fluid pressure that appears in \eqref{eq:eomtrad} via the stress tensor \cite{Armas:2018zbe}\footnote{Both $h^\mu$ and $\mu$ get corrections at high-derivative orders in terms of the magnetic fields and its derivatives. These corrections are frame dependent and have been explicitly calculated for the Landau frame in \cite{Armas:2018zbe} where $\mu$ here is identified with $\varpi$ in \cite{Armas:2018zbe}.}. Furthermore, the existence of a conserved two-form current $J^{\mu\nu}$ as in \eqref{eq:eomdual} implies the existence of a conserved charge 
\begin{equation}
Q=\int_{\mathcal{M}_2}\star J~~,
\end{equation}
where $\star$ is the Hodge operator in four-dimensional spacetime. In this context, this charge is interpreted as counting the number of magnetic field lines crossing a transverse two-dimensional surface $\mathcal{M}_2$ \cite{Gaiotto:2014kfa, Grozdanov:2016tdf}.

The main advantage of this formulation of MHD lies in the fact that recasting the Bianchi identity as a conservation law for a two-form current allows for a more systematic rearrangement of the hydrodynamic derivative expansion. Furthermore, in this formalism the constitutive relations for the conserved quantities $T^{\mu\nu}$ and $J^{\mu\nu}$ are solely written in terms of the temperature $T$ and quantities related to the magnetic field, without the necessity of introducing dependent degrees of freedom such as the electric field $E^\mu$. The electric fields can be easily obtained from \eqref{eq:Jdual}, in particular $E_\mu=-\epsilon_{\mu\nu\rho\sigma}u^\nu J^{\rho\sigma}$.

We stress that formally, we should view Eqs.~\eqref{eq:eomdual}, together with an additional Josephson condition, as equations describing a system with spontaneously broken one-form symmetry. While not necessary for the purposes of this work, we summarise this point and provide details about equilibrium partition functions for one-form hydrodynamics in App.~\ref{app:details1form}. Regardless, all that is left for having a well defined system with dynamics given by \eqref{eq:eomdual} is to provide explicit expressions for $T^{\mu\nu}$ and $J^{\mu\nu}$, as discussed below.

\subsection{Constitutive relations}
In order to specify the form of $T^{\mu\nu}$ and $J^{\mu\nu}$ in terms of the dynamical fields, we perform a gradient expansion such that 
\begin{equation}
T^{\mu\nu}=T^{\mu\nu}_{(0)}+T^{\mu\nu}_{(1)}+\mathcal{O}(\partial^2)~~,~~J^{\mu\nu}=J^{\mu\nu}_{(0)}+J^{\mu\nu}_{(1)}+\mathcal{O}(\partial^2)~~,
\end{equation}
where the subscript $(i)$ denotes the order of derivatives. In this paper we are only interested in first order corrections so we discard $\mathcal{O}(\partial^2)$ terms. At ideal order we have the following constitutive relations \footnote{Here the pressure $p$ is related to the pressure $P$ appearing in $\mu=-2|B|\partial P/\partial |B|^2$ via the relation $p(T,\mu)=P(T,B^2)-2|B|^2\partial P/\partial |B|^2$ \cite{Armas:2018atq}.}
\begin{equation}
T^{\mu\nu}_{(0)}=(\epsilon+p)\,u^\mu u^\nu+p\,g^{\mu\nu}-\mu\rho h^\mu h^\nu~~,~~ J^{\mu\nu}_{(0)}=2\rho\, u^{[\mu} h^{\nu]}~~,
\end{equation}
where $g_{\mu\nu}$ is the spacetime four dimensional metric, $\epsilon$ is the energy density, $p$ the pressure and $\rho$ the charge density. This charge density describes the density of objects charged under the one-form symmetry or, which in this context corresponds to the density of magnetic field lines. The parameters $\epsilon, p, \rho$ are all functions of $T,\mu$ and satisfy the thermodynamic identities
\begin{equation} \label{eq:GD}
\epsilon+p=Ts+\mu\rho~~,~~dp=sdT+\rho d\mu~~,    
\end{equation}
where $s$ is the entropy density. It is clear from Eq.~\eqref{eq:GD} that $\mu$ can be interpreted as the conjugate potential to the density $\rho$ of magnetic field lines. In particular, in addition to the equations of motion \eqref{eq:eomdual} we have assumed the existence of an entropy current $S^\mu=S^\mu_{(0)}+S^\mu_{(1)}+\mathcal{O}(\partial^2)$ with $S^\mu_{(0)}=s u^\mu$ that obeys a local version of the second law of thermodynamics, namely $\nabla_\mu S^\mu\ge0$.

In turn, at order $\mathcal{O}(\partial)$ the constitute relations can be parametrised as 
\begin{equation} \label{eq:stress1}
\begin{split}
&T^{\mu\nu}_{(1)}=\delta \varepsilon u^{\mu}u^{\nu}+\delta f \Delta^{\mu\nu}+\delta \tau h^{\mu}h^{\nu}+2\delta \chi h^{(\mu}u^{\nu)}+2\ell^{(\mu}h^{\nu)}+2k^{(\mu}u^{\nu)}+t^{\mu\nu}~~,\\
&J^{\mu\nu}_{(1)}=2\delta\varrho u^{[\mu}h^{\nu]}+2m^{[\mu}h^{\nu]}+2n^{[\mu}u^{\nu]}+s^{\mu\nu}~~,
\end{split}
\end{equation}
where $\Delta^{\mu\nu}=g^{\mu\nu}+u^{\mu}u^{\nu}-h^{\mu}h^{\nu}$ is a perpendicular projector to the subspace perpendicular to both $u^\mu$ and $h^\mu$, where the vectors $\ell^\mu, k^\mu, m^\mu, n^\mu$ and the traceless tensors  $t^{\mu\nu}=t^{(\mu\nu)}$ and $s^{\mu\nu}=s^{[\mu\nu]}$ are defined. Notice that the fact that these vectors are defined in the perpendicular space guarantees that the vector $h^\mu$ remains parallel to the magnetic field $B^\mu=J^{\mu\nu}u_\nu$ at first order in the gradient expansion. We now parametrise the various quantities in \eqref{eq:stress1} in terms of various transport coefficients and one-derivative tensor structures that are invariant under both parity and charge conjugation. This leads to 
\begin{equation} \label{eq:param}
\begin{split}
&\delta \varepsilon=-\varepsilon_1\Delta^{\mu\nu}\nabla_\mu u_\nu -\varepsilon_2 h^\mu h^\nu\nabla_\mu u_\nu-\varepsilon_3 u^\mu\nabla_\mu T-\varepsilon_4u^\mu\nabla_\mu\left( \mu/T\right)~~,\\
&\delta f=-f_1\Delta^{\mu\nu}\nabla_\mu u_\nu -f_2 h^\mu h^\nu\nabla_\mu u_\nu-f_3 u^\mu\nabla_\mu T-f_4u^\mu\nabla_\mu\left( \mu/T\right)~~,\\
&\delta \tau=-\tau_1\Delta^{\mu\nu}\nabla_\mu u_\nu -\tau_2 h^\mu h^\nu\nabla_\mu u_\nu-\tau_3 u^\mu\nabla_\mu T-\tau_4 u^\mu\nabla_\mu\left( \mu/T\right)~~,\\
&\delta\chi=-T\chi_1 u^\mu h^\nu\delta_B g_{\mu\nu}-\chi_2 \nabla_\mu(T\rho h^\mu)~~,\\
&\ell^\mu=-T\ell_1\Delta^{\mu\sigma}h^{\nu}\delta_B g_{\nu\sigma}-T\ell_2 \Delta^{\mu\sigma}u^{\nu}\delta_B b_{\sigma\nu}~~,\\
&k^\mu=-Tk_1\Delta^{\mu\nu}h^\lambda\delta_B b_{\nu\lambda}-Tk_2 \Delta^{\mu\nu}u^\lambda\delta_Bg_{\nu\lambda}\\
&\delta\varrho=-\varrho_1\Delta^{\mu\nu}\nabla_\mu u_\nu -\varrho_2 h^\mu h^\nu\nabla_\mu u_\nu-\varrho_3 u^\mu\nabla_\mu T-\varrho_4u^\mu\nabla_\mu\left( \mu/T\right)~~,\\
&m^\mu=-Tm_1\Delta^{\mu\nu}h^\lambda\delta_B b_{\nu\lambda}-Tm_2 \Delta^{\mu\nu}u^\lambda\delta_Bg_{\nu\lambda}~~, \\
&n^\mu=-Tn_1\Delta^{\mu\sigma}h^{\nu}\delta_B g_{\nu\sigma}-Tn_2\Delta^{\mu\sigma}u^{\nu}\delta_B b_{\sigma\nu}~~,\\
&t^{\mu\nu}=-T\eta_{\perp}\left(\Delta^{\mu\rho}\Delta^{\nu\sigma}-\frac{1}{2}\Delta^{\mu\nu}\Delta^{\rho\sigma}\right)\delta_Bg_{\rho\sigma}~~,\\
&s^{\mu\nu}=-T r_{||}\Delta^{\mu\rho}\Delta^{\nu\sigma}\delta_B b_{\rho\sigma}~~,
\end{split}
\end{equation}
where $\varepsilon_i,f_i,\tau_i,\chi_i,\ell_i,k_i,\varrho_i,m_i,n_i, \eta_\perp, r_{||}$ are arbitrary transport coefficients (and functions of $T,\mu$) and where we have defined
\begin{equation} \label{eq:deltaB}
\delta_B g_{\mu\nu}=2\nabla_{(\mu}\left(\frac{u_{\nu)}}{T}\right)~~,~~\delta_B b_{\mu\nu}=2\partial_{[\mu}\left(\frac{\mu h_{\nu]}}{T}\right)+\frac{u^\sigma}{T}H_{\sigma\mu\nu}~~.
\end{equation}
All the tensor structures appearing in \eqref{eq:param} vanish in equilibrium ($\delta_B g_{\mu\nu}=\delta_B b_{\mu\nu}=\nabla_\mu(T\rho h^\mu)=0$). This is because, under the assumption of parity and charge conjugation symmetry, the equilibrium partition function vanishes at first order in derivatives \cite{Armas:2018atq, Armas:2018zbe}.\footnote{See App.~\ref{app:discrete} for the transformation properties of the relevant tensor structures under parity and charge conjugation.} Thus, in a general frame, we are required to work with 28 transport coefficients at first order in derivatives. However, not all coefficients are genuine coefficients due to the freedom of performing frame transformations, which we discuss next.

\subsection{Frame transformations}
Out of equilibrium the fluid variables $T, \mu, u^\mu, h^\mu$ are not uniquely defined. It is always possible to redefine them by adding terms of order $\mathcal{O}\left(\partial\right)$ such that
\begin{equation}
T\to T+\delta T~~,~~\mu\to\mu+\delta\mu~~,~~u^\mu\to u^\mu+\delta u^\mu~~,~~h^\mu\to h^\mu+\delta h^\mu~~,    
\end{equation}
subject to the conditions
\begin{equation}
u_\mu\delta u^\mu=0~~,~~h_\mu\delta h^\mu=0~~,~~h_\mu\delta u^\mu=-u_\mu\delta h^\mu~~.
\end{equation}
These frame transformations lead to variations of the currents $\delta T^{\mu\nu}_{(1)}$ and $\delta J^{\mu\nu}_{(1)}$. It is possible to parameterize the most general frame one-derivative transformation according to 
\begin{equation}
    \delta u^\mu=\Delta^{\mu\nu}\alpha_\nu+\tilde\beta h^\mu~~,~~\delta h^\mu=\Delta^{\mu\nu}\gamma_\nu+\tilde\beta u^\mu~~,
\end{equation}
for some vectors $\alpha_\mu, \gamma_\mu$ and scalar $\tilde\beta$ respecting parity and charge conjugation symmetry such that
\begin{equation}
\begin{split}
&\gamma^\mu=-T\gamma_{1}\Delta^{\mu\nu}h^\sigma\delta_B g_{\nu\sigma}-T\gamma_{2}\Delta^{\mu\nu}u^\sigma \delta_B b_{\nu\sigma}~~,\\
&\alpha^\mu=-T\theta_1\Delta^{\mu\nu}h^\lambda\delta_B b_{\nu\lambda}-T\theta_2 \Delta^{\mu\nu}u^\lambda\delta_Bg_{\nu\lambda}~~,\\
&\tilde\beta=-T\tilde\gamma_1 u^\mu h^\nu\delta_B g_{\mu\nu}-\tilde\gamma_{2}\nabla_\mu(T\rho h^\mu)~~,\\
&\delta T=-t_1\Delta^{\mu\nu}\nabla_\mu u_\nu -t_2 h^\mu h^\nu\nabla_\mu u_\nu-t_3 u^\mu\nabla_\mu T-t_4u^\mu\nabla_\mu\left( \mu/T\right)~~,\\
&\delta \mu=-\omega_1\Delta^{\mu\nu}\nabla_\mu u_\nu -\omega_2 h^\mu h^\nu\nabla_\mu u_\nu-\omega_3 u^\mu\nabla_\mu T-\omega_4u^\mu\nabla_\mu\left( \mu/T\right)~~.
\end{split}
\end{equation}
This arbitrary change of frame transforms the coefficients appearing in Eq.~\eqref{eq:stress1} in the following manner
\begin{equation} \label{eq:framchange}
\begin{split}
&\ell_i\to\ell_i-\mu\rho \gamma_i~~,~~i=1,2\\
&n_i\to n_i-\rho \gamma_i~~,~~i=1,2\\
&m_i\to m_i+\rho \theta_i~~,~~i=1,2\\
&k_i\to k_i+(\epsilon+P)\theta_i~~,~~i=1,2\\
&\chi_i\to \chi_i+Ts \tilde\gamma_i~~,~~i=1,2\\
&\varepsilon_i\to\varepsilon_i+\frac{\partial\epsilon}{\partial T}t_i+\frac{\partial\epsilon}{\partial \mu}\omega_i~~,~~i=1,2,3,4\\
&f_i\to f_i+\frac{\partial P}{\partial T}t_i+\frac{\partial P}{\partial \mu}\omega_i~~,~~i=1,2,3,4\\
&\tau_i\to\tau_i+\left(\frac{\partial P}{\partial T}-\mu\frac{\partial\rho}{\partial T}\right)t_i-\mu\frac{\partial\rho}{\partial\mu} \omega_i~~,~~i=1,2,3,4\\
&\varrho_i\to\varrho_i+\frac{\partial\rho}{\partial T}t_i+\frac{\partial \rho}{\partial \mu}\omega_i~~,~~i=1,2,3,4~~,\\
&\eta_\perp\to\eta_\perp~~,~~r_{||}\to r_{||}~~.
 \end{split}  
\end{equation}
Clearly, the last two coefficients are frame invariant but the others are not. It is possible to find frame invariant combinations besides $\eta_\perp,r_{||}$. These are given by the combinations 
\begin{equation} \label{eq:frameinvariant}
\begin{split}
&\tilde m_i\equiv m_i-\frac{\rho}{\epsilon+P}k_i~~,~~i=1,2\\
&\tilde \ell_i\equiv \ell_i-\mu n_i~~,~~i=1,2\\
&\tilde f_i\equiv f_i-\frac{\partial P}{\partial \epsilon}|_{\rho}~\varepsilon_i-\frac{\partial P}{\partial \rho}|_{\epsilon}~\varrho_i~~,~~i=1,2,3,4\\
&\tilde \tau_i\equiv \tau_i-\frac{\partial \left(P-\mu\rho\right)}{\partial \epsilon}|_{\rho}~\varepsilon_i-\frac{\partial \left(P-\mu\rho\right)}{\partial \rho}|_{\epsilon}~\varrho_i~~,~~i=1,2,3,4~~.
 \end{split}  
\end{equation}
Thus, there is a total of 14 genuine frame-invariant transport coefficients. A common choice of frame is the Landau frame in which $\delta\varepsilon=\delta\chi=k^\mu=0$ and $\delta\varrho=n^\mu=0$. This can be accomplished by just setting the coefficients $\varepsilon_i,\chi_i,k_i,\varrho_i,n_i$ to zero or by a particular frame transformation that we specify in App.~\ref{app:Landauframe}.

\subsection{Entropy inequalities and microscopic CPT}
The coefficients appearing above are constrained by entropy production and microscopic CPT symmetry. In order to address the former, we consider the canonical entropy current
\begin{equation}
S^\mu=s u^\mu-\frac{1}{T}T^{\mu\nu}_{(1)}u_\nu-\frac{\mu}{T}J^{\mu\nu}_{(1)}h_\nu~~,
\end{equation}
and demand $\nabla_\mu S^\mu\ge0$ for all physical configurations. By \emph{physical configurations} we mean fluid configurations that satisfy the equations of motion \eqref{eq:eomdual}. By doing so we obtain the same constraints as those found in \cite{Armas:2018ibg, Armas:2018atq, Armas:2018zbe} but in terms of specific combinations of transport coefficients, in particular
\begin{equation}
    \label{S_constr}
    \boldsymbol{\zeta}_\perp,\boldsymbol{\zeta}_{||},\boldsymbol{r}_\perp,\boldsymbol{\eta}_\perp,\boldsymbol{\eta}_{||},\boldsymbol{r}_{||},\boldsymbol{r}_{\perp}\ge0  ~~,~~
    \boldsymbol{\zeta}_{\perp}\boldsymbol{\zeta}_{||}\ge\frac{1}{4}\left(\boldsymbol{\zeta}_{\times}+\boldsymbol{\zeta}_{\times}^{'}\right)^2~~,
\end{equation}
where we have defined 
\begin{equation} \label{eq:boldcoef}
\begin{split}
&\boldsymbol{\eta}_\perp=\eta_\perp~~,~~ \boldsymbol{r}_{||}=r_{||}~~,~~\boldsymbol \eta_{||}=\tilde \ell_1-\mu\tilde \ell_2~~,~~\boldsymbol{r}_{\perp}=\tilde m_1-\frac{\rho}{\epsilon+P}\tilde m_2~~,\\
&\boldsymbol{\zeta}_{\perp}=\tilde f_1-T\frac{\partial P}{\partial \epsilon}|_{\rho} ~\tilde f_3-\frac{1}{T}\frac{\partial P}{\partial \rho}|_{\epsilon}~\tilde f_4~~,\\
&\boldsymbol{\zeta}_{\times}=\tilde f_2-T\frac{\left(P-\mu\rho\right)}{\partial \epsilon}|_{\rho}~\tilde f_3-\frac{1}{T}\left(\frac{\left(P-\mu\rho\right)}{\partial \rho}|_{\epsilon}+\mu\right)\tilde f_4 ~~,\\
&\boldsymbol{\zeta}_{\times}^{'}=\tilde\tau_1-T \frac{\partial P}{\partial \epsilon}|_{\rho} \tilde\tau_3-\frac{1}{T}\frac{\partial P}{\partial \rho}|_{\epsilon}\tilde \tau_4~~,~~\boldsymbol{\zeta}_{||}=\tilde\tau_2-T\frac{\left(P-\mu\rho\right)}{\partial \epsilon}|_{\rho} \tilde\tau_3-\frac{1}{T}\left(\frac{\left(P-\mu\rho\right)}{\partial \rho}|_{\epsilon}+\mu\right)\tilde\tau_4~~.
\end{split}
\end{equation}
Each of these coefficients (in bold) is built out of the frame invariants \eqref{eq:frameinvariant}. The entropy current imposes inequalities in 8 transport coefficients but does not constrain the remaining ones. 

To understand the constraints imposed by microscopic CPT invariance we compute the Kubo formulae by perturbing the background sources $g_{\mu\nu}$ and $b_{\mu\nu}$ with plane wave profiles with non-zero frequency $\omega$ and vanishing momentum. Defining
\begin{equation}
\delta \mathbb{T}^{\mu\nu}=\frac{1}{2}G_{TT}^{\mu\nu,\lambda\rho}\delta g_{\lambda\rho}+\frac{1}{2}G_{TJ}^{\mu\nu,\lambda\rho}\delta b_{\lambda\rho}~~,~~\delta \mathbb{J}^{\mu\nu}=\frac{1}{2}G_{JT}^{\mu\nu,\lambda\rho}\delta g_{\lambda\rho}+\frac{1}{2}G_{JJ}^{\mu\nu,\lambda\rho}\delta b_{\lambda\rho}~~,
\end{equation}
where we have introduced the one point functions $\mathbb{T}^{\mu\nu}=\sqrt{-g}\langle T^{\mu\nu}\rangle$ and $\mathbb{J}^{\mu\nu}=\sqrt{-g}\langle J^{\mu\nu}\rangle$ (see \cite{Armas:2018zbe}), we explicitly extract the non-zero Green's functions
\begin{equation}\label{eq:Kubo}
    \begin{split}
        &\lim_{\omega\to0}\left(\frac{G_{TT}^{ii,ii}}{-i \omega}\right)=\boldsymbol{\eta}_\perp +\boldsymbol{\zeta}_\perp~~,~~\lim_{\omega\to0}\left(\frac{G_{TT}^{ii,zz}}{-i \omega}\right)=\boldsymbol{\zeta}_\times~~,~~\lim_{\omega\to0}\left(\frac{G_{TT}^{zz,ii}}{-i \omega}\right)=\boldsymbol{\zeta}_\times^{'}~~,
        \\
        &\lim_{\omega\to0}\left(\frac{G_{TT}^{zz,zz}}{-i \omega}\right)=\boldsymbol{\zeta}_{||}~~,~~\lim_{\omega\to0}\left(\frac{G_{TT}^{ii,jj}-G_{TT}^{ii,ii}}{2(-i\,\omega)}\right)=-\boldsymbol{\eta}_\perp~~,~~\lim_{\omega\to0}\left(\frac{G_{TT}^{xy,xy}}{-i \omega}\right)=\boldsymbol{\eta}_\perp~~,
        \\
        &\lim_{\omega\to0}\left(\frac{G_{TT}^{iz,iz}}{-i \omega}\right)=\boldsymbol{\eta}_{||}~~,~~\lim_{\omega\to0}\left(\frac{G_{JJ}^{xy,xy}}{-i \omega}\right)= \boldsymbol{r}_\parallel~~,~~\lim_{\omega\to0}\left(\frac{G_{JJ}^{iz,iz}}{-i \omega}\right)=\boldsymbol{r}_\perp~~,
    \end{split}
\end{equation}
where we have introduced flat cartesian coordinates $(t,x,y,z)$ and the index $i=x,y$. Given these Kubo formulae, Onsager's relations require that
\begin{equation} \label{eq:CPT}
\boldsymbol{\zeta}_\times=\boldsymbol{\zeta}_\times^{'}~~.
\end{equation}
As can be noticed, only the combinations of transport coefficients appearing in \eqref{eq:boldcoef} enter the Kubo formulae. Therefore, only these coefficients are strictly hydrodynamic, while other coefficients, which will be required to make the formulation of MHD presented here stable and causal are non-hydrodynamic. As we will show in the next section, several of the coefficients will have to satisfy certain constrains due to causality and stability of the equilibrium state.

%
%
%
%

\section{Constraints from stability and causality} \label{sec:constraints}
In this section we find the necessary conditions on the various transport coefficients introduced in \eqref{eq:param} in order for the system of equations \eqref{eq:eomdual} to yield a causal and stable evolution for  linear perturbations of a generic equilibrium state. We begin by defining the equilibrium state, the different classes of perturbations that we consider and the criteria we employ for stability and causality. In the next section we present a detailed analysis of the constrains implied by these criteria on the several transport coefficients for a particular plasma. 

\subsection{Equilibrium state, perturbations, causality and stability} \label{sec:equilibrium}
Equilibrium states of MHD have been extensively discussed in \cite{Armas:2018ibg, Armas:2018atq, Armas:2018zbe}. The most general equilibrium state is characterised by the vanishing of \eqref{eq:deltaB} together with the no-monopole constraint
\begin{equation} \label{eq:conditions}
\delta_B g_{\mu\nu}=0~~,~~\delta_B b_{\mu\nu}=0~~,~~\nabla_\mu\left(T\rho h^\mu\right)=0~~.   
\end{equation}
The first condition above implies that such states are characterised by the existence of a Killing vector field $K^\mu$ such that $u^\mu=K^\mu/|K|$ where $|K|=|-g_{\mu\nu}K^\mu K^\nu|^{1/2}$. In turn the temperature is given by $T=T_0/|K|$ where $T_0$ is a constant. The second and third conditions in \eqref{eq:conditions} are non-trivial and express the degeneracy of the equilibrium state very akin to the degeneracy encountered in the context of superfluids. In the present setting, this degeneracy expresses the fact that there are many possible magnetic field configurations in thermal equilibrium.\footnote{From a more formal point of view, this degeneracy is rooted in the possible equilibrium profiles for the magnetic scalar potential $\varphi$ that we introduce in App. \ref{app:details1form}.} We focus on flat backgrounds without external currents, in particular
\begin{equation}
g_{\mu\nu}=\eta_{\mu\nu}~~,~~b_{\mu\nu}=0~~,    
\end{equation}
with coordinates $(t,x,y,z)$ and we take the fluid to be at rest $u^\mu=(1,0,0,0)$ as well as thermodynamic parameters to be constant (e.g. $T=T_0$, $\mu=\mu_0$). In addition, we take $h^\mu=(0,0,0,1)$. This is the most generic equilibrium in which $h^\mu$ is aligned with a single spatial isometry of the background (in this case flat spacetime). It is possible to find other equilibrium configurations for which the direction of the magnetic field $h^\mu$ is not aligned with a spatial isometry (e.g. rotating magnetised plasmas/stars \cite{Armas:2018atq}). Such configurations involve fluid velocities and magnetic fields that depend on the spacetime coordinates but which locally are described by the equilibrium configurations that we consider here. It would be interesting to consider such configurations but for simplicity we take $h^\mu=(0,0,0,1)$.\footnote{This is similar to the approach taken in \cite{Kovtun:2019, Hoult:2020eho} in which the analysis assumes equilibrium configurations with constant fluid velocities.} We also discuss equilibrium states at finite velocity $\boldsymbol{\beta}$, which can be obtained by performing a Lorentz boost (see App.~\ref{app:boostedframe}). In addition, the equilibrium states considered here satisfy the following inequalities
\begin{equation} \label{eq:assthermal}
    \begin{split}
        &(\epsilon+p)>0~~,~~\mu\rho>0~~,~~s>0~~,~~T>0~~,\\
        &\left(\frac{\partial \rho}{\partial \mu}\right)_T=\chi\geq0~~,
        \\
        &\frac{V^2}{\chi}=T\left(\frac{\partial \epsilon}{\partial T}\right)_\mu+\mu\left(\frac{\partial \epsilon}{\partial \mu}\right)_T=cT^2+2\mu T \lambda+\mu^2\chi\geq0~~,
        \\
        &\frac{U^2}{T}=\left(\frac{\partial \epsilon}{\partial T}\right)_\mu\left(\frac{\partial \rho}{\partial \mu}\right)_T-\left(\frac{\partial \rho}{\partial T}\right)_\mu\left(\frac{\partial \epsilon}{\partial \mu}\right)_T=T(c\chi-\lambda^2)\geq0~~,
    \end{split}
\end{equation}
where we have dropped the subscript $0$ in all equilibrium thermodynamic quantities for simplicity and defined
\begin{equation} \label{eq:equalities}
c=\left(\frac{\partial s}{\partial T}\right)_\mu ~~,~~\lambda=\left(\frac{\partial s}{\partial \mu}\right)_T=\left(\frac{\partial \rho}{\partial T}\right)_\mu~~.
\end{equation}
The last set of equalities in \eqref{eq:equalities} arises from the fact that both the entropy $s$ and the "charge density" $\rho$ in equilibrium are determined in terms of the pressure $p$. The last three inequalities in \eqref{eq:assthermal} can be derived from general thermodynamic considerations (see App.~\ref{app:details1form}). However, such inequalities can also be obtained by studying the constraints arising from stability and causality of the excitations of the system, as we will comment throughout. 

Given the equilibrium state, we subsequently consider linear perturbations around it of the plane wave form. Specifically
\begin{equation}
u^\mu\rightarrow u^\mu+  e^{i k\cdot x}\,\delta u^\mu~~,~~h^\mu\rightarrow h^\mu+  e^{i k\cdot x}\,\delta h^\mu~~,~~\mu\rightarrow \mu+ e^{i k\cdot x}\,\delta \mu~~,~~T\rightarrow T+  e^{i k\cdot x}\,\delta T~~,
\end{equation}
where we have parameterised the four wave vector as $k^\mu=(\omega, k\sin\theta, 0, k\cos\theta)$ for frequency $\omega$ and wavenumber $k$ and where $\theta$ is an angle between the direction of the background magnetic field and the waves' momentum. Introducing these perturbations in \eqref{eq:eomdual} leads to a characteristic equation of the form $F(\omega,k,\boldsymbol{\beta})=0$ whose solutions $\omega(k,\boldsymbol{\beta})$ are the excitations/modes of the system. These perturbations can be decomposed into two channels. The Alfvén channel is defined via the constraints on perturbations
\begin{equation} \label{A_cond}
\delta \mu =\delta T=0~~,~~ u_\mu \delta h^\mu=0~\Rightarrow h_\mu \delta u^\mu=0~~,~~k_\mu \delta u^\mu=k_\mu \delta h^\mu=0~~.
\end{equation}
In this case the characteristic equation $F(\omega,k,\boldsymbol{\beta})=0$ yields 2 gapless modes and 2 gapped modes (i.e. modes for which $\omega(k\to0)\ne0$). In turn the magnetosonic channel is defined by non-vanishing $\delta T$ and $\delta \mu$ and by amplitudes $\delta u^\mu$ and $\delta h^\mu$ that lie in the subspace spanned by $\{u^\mu,\,h^\mu,\,k^\mu\}$ (see e.g. \cite{Grozdanov:2016tdf, Grozdanov:2017kyl}). In this channel, we find 4 gapless modes and 5 gapped modes at low wavenumber $k$.

Our goal in this work is to study the spectrum of excitations arising from these linear perturbations and make sure that the transport coefficients appearing in \eqref{eq:param} are constrained in such a way as not to lead to instabilities of the equilibrium state or to violations of causality. To that aim, we adopt the same criteria as in \cite{Kovtun:2019, Hoult:2020eho}, in particular
\begin{equation} \label{eq:criteria}
\text{Im}\left(\omega(k)\right)<0~~,~~1>\lim_{k\to\infty}\left| \frac{\text{Re}\left(\omega(k)\right)}{k}\right|>0~~.    
\end{equation}
The first condition in \eqref{eq:criteria} ensures stability of the perturbations while the second condition ensures causality (i.e. signals do not propagate faster than the speed of light).\footnote{See \cite{cmp/1103904078} for a discussion of causality criteria.} It should be noted that the second condition involves the limit $k\to\infty$ and the first condition must be imposed for all $k$. In other words, we are aiming at having a well defined evolution even for very short wavelengths, that is, beyond the traditional regime of validity of hydrodynamics (i.e. the small $k$ regime).
In practice, the constrains can be quite unwieldy so the strategy we adopt here is to first study the constraints in the small and large $k$ regimes and then use those constraints to simplify the analysis at arbitrary $k$. We approach both channels separately. 

\subsection{Constraints in the Alfvén channel}

We wish to find a fluid frame that is stable and causal for any boost parameter $\boldsymbol{\beta}$. However, in the present case explicit results for any $\boldsymbol{\beta}$ are cumbersome. In addition, Ref.~\cite{Bemfica:2020zjp} has established a theorem stating that\footnote{See sec. VI of \cite{Bemfica:2020zjp} for an explicit statement of the theorem.}, for isotropic fluids without magnetic fields, whenever the linearised equations of motion around an equilibrium state with $\boldsymbol{\beta}=0$ are stable and causal, they remain so for $\boldsymbol{\beta}\ne0$.\footnote{Notice that these conditions are not met for the Landau frame, as was shown for instance in the case of an uncharged fluid in ref. \cite{Kovtun:2019}. The Landau frame with $\boldsymbol{\beta}=0$ is stable but not causal.} Even though we have not proven this in the context of MHD, we expect it to hold and we explicitly verify this to be the case for the specific plasma of Sec.~\ref{sec:EOS}. Hence, we can limit our considerations to the rest frame, as this is sufficient to ensure stability and causality for $\boldsymbol{\beta}\ne0$ in both channels. We thus begin by studying the small $k$ regime for $\boldsymbol{\beta}=0$. In this case, the characteristic equation leads to two gapless modes and two gapped modes of the form
\begin{equation}\label{eq:wlowkAl}
\begin{split}
    &\omega=\pm \mathcal{V}_A k\cos\theta - \frac{i}{2} \left(\frac{\bs{\mathcal{T}_\eta}}{\epsilon+p}+\bs{\mathcal{T}_r}\frac{\mu}{\rho}\right)k^2 +\mathcal{O}(k^3)~~,
    \\ 
    &\omega=i\frac{\rho}{\mu\,n_2}+\mathcal{O}(k^2)~~,~~\omega=i\frac{(\epsilon+p)}{k_2}+\mathcal{O}(k^2)~~,
\end{split}
\end{equation}
where we have defined 
\begin{equation}
\begin{split}
&\mathcal{V}_A^2=\frac{\mu\rho}{\epsilon+p}~~,~~\boldsymbol{\mathcal{T}_\eta}=\boldsymbol{\eta}_{||}\cos^2\theta+\boldsymbol{\eta}_\perp\sin^2\theta~~,~~\boldsymbol{\mathcal{T}}_r=\boldsymbol{r}_\perp \cos^2\theta+\boldsymbol{r}_{||}\sin^2\theta~~.
\end{split}
\end{equation}
In particular $\mathcal{V}_A$ denotes the velocity of Alfvén waves. At ideal order, stability implies that $\mathcal{V}_A^2>0$ which is compatible with \eqref{eq:assthermal}. The form of the gapped modes in \eqref{eq:wlowkAl} implies that in the Landau frame, in which $n_2=k_2=0$, the criteria for stability in \eqref{eq:criteria} leads to instabilities of the equilibrium state. As such, the Landau frame, similarly to previous cases studied in \cite{Kovtun:2019}, is not the appropriate frame for a well defined initial value problem. From \eqref{eq:wlowkAl}, \eqref{eq:assthermal} and the criteria \eqref{eq:criteria} one concludes that the necessary conditions for the gapped modes to be stable are
\begin{equation} \label{eq:cc1}
    k_2<0~~, \quad n_2<0~~.
\end{equation}
The form of the gapless modes in \eqref{eq:wlowkAl} agrees with that of \cite{Grozdanov:2016tdf} in the Landau frame, and their stability in the hydrodynamic regime is guaranteed by the fact that $\boldsymbol{\mathcal{T}}_r>0$ and $\boldsymbol{\mathcal{T}}_\eta>0$ due to the entropy constraints \eqref{S_constr} and the conditions \eqref{eq:assthermal} on the equilibrium state. 

Let us now study the fluid at rest but at arbitrary wave number $k$. To this aim we consider the quantity $\Delta=i\omega$, determined by the zero of the characteristic function
\begin{equation}
\begin{split}
    F(\Delta,\bs{k},\bs{\beta}=0)=&\left[k^2\mathcal{T}_\eta
+\Delta(\epsilon+p-k_2\Delta)\right]\left[\mu k^2\mathcal{T}_r+\Delta(\rho-n_2\Delta \mu)\right]\\
&+k^2 \mu\left[\rho-(k_1+\ell_2)\Delta\right]\left[\rho-(m_2+n_1)\Delta\right]\cos^2\theta~~,
\end{split}
\end{equation}
where we have defined
\begin{equation}
    \mathcal{T}_\eta=\ell_1\cos^2\theta+\eta_\perp\sin^2\theta~~,~~\mathcal{T}_r=m_1 \cos^2\theta+r_{||}\sin^2\theta~~.
\end{equation}
More specifically, $\Delta$ is the solution of a quartic algebraic equation $ \Delta^4+ A^{(k)}_3\Delta^3+A^{(k)}_2\Delta^2+A^{(k)}_1\Delta+A^{(k)}_0=0$. We record the explicit expression for the coefficients
\begin{equation} \label{Alfvén_poly}
    \begin{split}
        A^{(k)}_3&=-\left(\frac{\rho}{n_2\mu}+\frac{\mu\rho}{\mathcal{V}_A^2 k_2}\right)~~,
        \\
        A^{(k)}_2&=\frac{\rho}{n_2\mu}\frac{\mu\rho}{\mathcal{V}_A^2 k_2}-k^2\left[\frac{\mathcal{T}_r}{n_2}+\frac{\mathcal{T}_\eta}{k_2}-\frac{(k_1+\ell_2)(m_2+n_1)}{k_2n_2}\cos^2\theta\right]
        \\
        A^{(k)}_1&=k^2\left[\frac{\mathcal{T}_r}{n_2}\frac{\mu\rho}{\mathcal{V}_A^2 k_2}+\frac{\mathcal{T}_\eta}{k_2}\frac{\rho}{n_2\mu}-\rho\frac{(k_1+\ell_2+m_2+n_1)}{k_2 n_2} \cos^2\theta\right]~~,
        \\
        A^{(k)}_0&=\frac{\rho}{n_2\mu}\frac{\mu\rho}{\mathcal{V}_A^2 k_2}k^2\mathcal{V}_A^2\cos^2\theta+k^4\frac{\mathcal{T}_r}{n_2}\frac{ \mathcal{T}_\eta}{k_2}~~.
    \end{split}
\end{equation}
We have to study stability and causality for this polynomial for $k>0$. The stability conditions follow from the Routh-Hurwitz theorem, which for the polynomial at hand amounts to demand
\begin{equation}
\label{eq: Alfv_stability}
    A^{(k)}_3>0~,~ A^{(k)}_0>0~,~\quad \left(A^{(k)}_3 A^{(k)}_2- A^{(k)}_1\right)A^{(k)}_1-\left(A^{(k)}_3\right)^2A^{(k)}_0>0~~.
\end{equation}
Even though eq. \eqref{eq:cc1} presents necessary conditions for the modes to be stable, stability must be imposed at arbitrary values of $k$, as well as for every angle $\theta$. From the conditions in \eqref{eq: Alfv_stability} we see that the first is already ensured by \eqref{eq:cc1} and \eqref{eq:assthermal} while the second implies that 
\begin{equation}
    \mathcal{T}_r\mathcal{T}_\eta>0~~\Rightarrow~~ \ell_1 m_1>0~~,~~\ell_1 r_{||}+\eta_\perp m_1>0~~,
\end{equation}
where we have used the entropy constraints $\eta_\perp r_{||}>0$. It is unwieldy to deduce constraints on the last condition in \eqref{eq: Alfv_stability} but a necessary condition is that $A^{(k)}_1>0$ which can be satisfied by imposing $\mathcal{T}_r>0$ and $\mathcal{T}_\eta>0$ in turn implying $m_1,\ell_1>0$.

The causality conditions applies at short-wavelengths $k\to\infty$ according to eq. \eqref{eq:criteria}. In this limit the dispersion relations becomes linear $\omega=W k+\mathcal{O}(k^0)$ and
$W^2$ is determined by the second order equation $W^4+A_2^{(\infty)}W^2+A_0^{(\infty)}=0$, with 
\begin{equation}
    A_2^{(\infty)}=\frac{\mathcal{T}_r}{n_2}+\frac{\mathcal{T}_\eta}{k_2}-\frac{(k_1+\ell_2)(m_2+n_1)}{k_2n_2}\cos^2\theta~,~
    A^{(\infty)}_0=\frac{\mathcal{T}_r}{n_2}\frac{ \mathcal{T}_\eta}{k_2}~~.
\end{equation}
Given that $W$ is the phase velocity, one must have $1>W^2\geq0$, which follows after imposing
\begin{equation}
\label{eq: Alfv_causality}
    \left(A^{(\infty)}_2\right)^2-4A^{(\infty)}_0>0~,~ A^{(\infty)}_2<0~,~ A^{(\infty)}_0+A^{(\infty)}_2+1>0~~.
\end{equation}
The first of these conditions ensures that $W$ is real, and thus $W^2>0$, whereas the other two conditions are meant to prevent superluminal propagation of Alfvén modes. It is possible to solve the conditions \eqref{eq: Alfv_causality} by a suitable choice of frame. In particular choosing $k_1=-\ell_2$ or $m_2=-n_1$ eliminates the dependence on $\theta$ from $A_2^{(\infty)}$. The inequalities \eqref{eq: Alfv_causality} can then
be solved by requiring
\begin{equation} \label{eq:cond2}
   \mathcal{T}_r>0~~,~~\mathcal{T}_\eta>0~~,~~\mathcal{T}_r k_2+n_2\mathcal{T}_\eta+\mathcal{T}_r\mathcal{T}_\eta+n_2k_2>0~~,
\end{equation}
which in turn lead to sufficient conditions
\begin{equation}
m_1\ell_1+k_2m_1+\ell_1 n_2>0~~,~~\eta_\perp r_{||}+k_2r_{||}+n_2\eta_\perp>0~~,~~\eta_\perp(m_1-r_{||})+\ell_1(r_{||}-m_1)>0~~.
\end{equation}
We note that even though the conditions \eqref{eq:cond2} were derived for the specific frame $k_1=-\ell_2$ or $m_2=-n_1$, they are actually necessary conditions since causality at $\theta=\pi/2$ enforces them.

To summarize, in the Alfvén channel the stability conditions at arbitrary $k$ and $\theta$ and for a generic equation of state are given by eqs. \eqref{eq:cc1} and \eqref{eq: Alfv_stability} while the causality conditions at arbitrary $\theta$ and for a generic equation of state are given in eq. \eqref{eq: Alfv_causality}.

\subsection{Constraints in the magnetosonic channel} \label{sec:cM}

As in the Alfvén channel, we study the system in the rest frame $\boldsymbol{\beta}=0$ and at small $\boldsymbol{k}$. In this limit one finds four gapless modes, corresponding to the polarizations of the fast and slow magnetosonic waves
\begin{equation}
    \omega=\pm v_M \, k-\frac{i}{2}\mathcal{\tau} k^2+\mathcal{O}(k^3)~~,
\end{equation}
where the phase velocity reads
\begin{equation} \label{eq:gaplessmag}
    v_M^2=\frac{1}{2}\left[(\mathcal{V}_A^2+\mathcal{V}_0^2)\cos^2\theta+\mathcal{V}_s^2\sin^2\theta\right]\pm\frac{1}{2}\sqrt{\left[(\mathcal{V}_A^2-\mathcal{V}_0^2)\cos^2\theta+\mathcal{V}_s^2\sin^2\theta\right]^2+4\mathcal{V}^4\cos^2\theta\sin^2\theta}~~,
\end{equation}
and where we have defined
\begin{equation}
    \label{VVV}
    \mathcal{V}_0^2=\frac{s \chi}{T(c \chi - \lambda^2)}~~,~~ \mathcal{V}_s^2=\frac{s^2\chi+c\rho^2-2s\rho \lambda}{(\epsilon+p)(c\chi-\lambda^2)}~~,~~ \mathcal{V}^4=\frac{s(\rho \lambda-s \chi)^2}{T(c \chi-\lambda^2)^2(\epsilon+p)}~~.
\end{equation}
The expression \eqref{eq:gaplessmag} agrees with that obtained in \cite{Grozdanov:2016tdf} in the Landau frame. Reality of the phase velocity implies $\mathcal{V}_0^2>0$ and $\mathcal{V}_s^2>0$ which is consistent with the thermodynamic assumptions \eqref{eq:assthermal}. In addition, $\mathcal{V}^4>0$ is implied by \eqref{eq:assthermal}. The signs appearing in \eqref{eq:gaplessmag} characterise the fast ($+$) and slow ($-$) magnetosonic modes. The diffusion rate $\tau$ satisfies $\tau>0$ given the entropy constraints \eqref{S_constr}. The expression for $\tau$ is lengthy but it simplifies for specific choices of angles, e.g. 
\begin{equation}
\begin{split}
    &\tau(v_M=0,\theta=\pi/2)=\frac{1}{sT}\left[\boldsymbol{\eta}_{\parallel}+\frac{(\epsilon+p)\mathcal{V}_0^2}{\chi\mathcal{V}_s^2}\boldsymbol{r}_\perp\right]~~,
    \\
    &\tau(v_M=\mathcal{V}_s,\theta=\pi/2)=\frac{1}{sT}\left[(1-\mathcal{V}_A^2)(\boldsymbol{\eta}_{\perp}+\boldsymbol{\zeta}_\perp)+\frac{\mathcal{V}_0^2}{\chi\mathcal{V}_s^2}[(\epsilon+p)-\mathcal{V}_s^2(cT^2+2\mu T \lambda +\mu^2\chi)]\boldsymbol{r}_\perp\right]~~,
    \\
    &\tau(v_M=\mathcal{V}_A,\theta=0)=\frac{(1-\mathcal{V}_A^2)}{sT}\boldsymbol{\eta}_{\parallel}+\frac{\mu}{\rho}\boldsymbol{r}_\perp~~,
    \\
    &\tau(v_M=\mathcal{V}_0,\theta=0)=\frac{1}{sT}\boldsymbol{\zeta}_{\parallel}~~.
\end{split}
\end{equation}
Additionally, there are five different gapped modes, which read
\begin{equation}\label{eq:gappedmag}
\begin{split}
    \omega&=i\frac{\rho}{\mu n_2}+\mathcal{O}(k^2)~~,~~\omega=i \frac{\epsilon+p}{k_2}+\mathcal{O}(k^2)~~,
    \\
    \omega&=i\frac{s T}{\chi_1+\chi_2 \rho T}+\mathcal{O}(k^2)~~,
    ~~\omega=i\,\frac{g\pm \sqrt{g^2+4T^4(\varepsilon_4\varrho_3-\varepsilon_3\varrho_4)(c\chi-\lambda^2)}}{2T(\varepsilon_4\varrho_3-\varepsilon_3\varrho_4)}+\mathcal{O}(k^2)~~,
\end{split}
\end{equation}
where we have defined $g=T \lambda(T^2\varrho_3+\varepsilon_4-2\mu\varrho_4)-T^2\chi(\varepsilon_3-\mu\varrho_3)+\mu \chi(\varepsilon_4-\mu\varrho_4)-cT^2\varrho_4$. Note that, as in the Alfvén channel, \eqref{eq:gappedmag} predicts that the equilibrium state is unstable in the Landau frame. Stability of the gapped modes requires that the $\mathcal{O}(k^0)$ terms in \eqref{eq:gappedmag} are strictly negative. This implies that we must have
\begin{equation}
\label{eq: MS_gaps_stability}
n_2<0~~,~~k_2<0~~,~~(\chi_1+\chi_2 \rho T)<0~~,~~g>0~~,~~(\varepsilon_4\varrho_3-\varepsilon_3\varrho_4)<0~~.    
\end{equation}
A general investigation of stability and causality in this channel is more involved compared to the Alfvén channel. The spectral function is in general a $9$th-degree polynomial for $\Delta=i\omega$ of the kind
\begin{equation}
    \label{Poly_P9}
    P_9(\Delta)=\Delta^9+\sum_{n=0}^{8}B_n^{(k)}\Delta^ n=0~~.
\end{equation}
The expressions for the coefficients are exceedingly cumbersome, and in general they are combination of ratios among the transport parameters and thermodynamics functions, together with  trigonometric functions of the angle $\theta$ and organised in powers in the momentum $k$. The stability conditions arising from this polynomial are detailed in app.~\ref{app:furtherdetails}. In general, as the polynomial does not factorise at arbitrary angle, an analytic analysis of the constraints is out of reach and a numerical analysis is necessary. 

Causality must be imposed on this channel. Similarly to the Alfvén channel, in the small wavelength regime $k\to\infty$ the dispersion relations becomes linear $\omega=W k+\mathcal{O}(k^0)$, and the polynomial $P_9(\Delta)$ in eq. \eqref{Poly_P9} reduces to a $4$th-degree polynomial in $W^2$
\begin{equation} \label{eq:Polycausal}
    P^{(\infty)}_4(W^2)=W^8+B^{(\infty)}_7 W^6+B^{(\infty)}_5 W^4+B^{(\infty)}_3 W^2+B^{(\infty)}_1,
\end{equation}
where the coefficient $B^{(\infty)}_n$ is defined as the leading order coefficient in a large $k$ expansion of the generic coefficient $B^{(k)}_n$ in eq. \eqref{Poly_P9}. In order to derive causality conditions we ensure Schur stability, which is detailed in app.~\ref{app:furtherdetails}. The conditions arising from this polynomial must be satisfied for every angle $\theta$.

To summarize, in the magnetosonic channel the stability and causality conditions at arbitrary $k$ and $\theta$ and for a generic equation of state are given by eqs. \eqref{eq: MS_gaps_stability} together with the general expressions provided in app.~\ref{app:furtherdetails}, in particular conditions \eqref{eq: MS_k_stability}, \eqref{eq: MS_Schur_causality} and \eqref{eq: MS_real_causality}. We will now study the constraints in detail in order to find a stable and causal class of frames for a specific plasma.

\section{Analysis of a specific equation of state}\label{sec:EOS}
Several of the constraints provided in the previous section are difficult to solve analytically due to the sheer number of coefficients and lengthy expressions. As in previous works \cite{Bemfica:2017wps, Bemfica:2019knx, Kovtun:2019, Bemfica:2020zjp, Hoult:2020eho}, it is possible to simplify the polynomials \eqref{Poly_P9} and \eqref{eq:Polycausal} by making a convenient choice of frame. In particular we choose
\begin{equation}
    \label{MS_choice1}
    \begin{split}
        \varepsilon_4=\varrho_3=\chi_2=0~~,
    \end{split}
\end{equation}
for which the stability conditions \eqref{eq: MS_gaps_stability} reduce to
\begin{equation} \label{eq:cc5}
n_2<0~~,~~k_2<0~~,~~\chi_1<0~~,~~\varrho_4<0~~,~~\varepsilon_3<0~~.
\end{equation}
In app.~\ref{app:furtherdetails} we give explicit forms for the polynomials which factorise for $\theta=0,\pi/2$ for the choice \eqref{MS_choice1}. However, requiring stability and causality in the magnetosonic channel for $\theta=0,\pi/2$ does not guarantee stability and causality at arbitrary angle $\theta$. In this section, as a proof of principle, we focus on a specific equation of state and perform a numerical analysis in order to identify a class of frames for which the spectrum is stable and causal.

\subsection{Equation of state and transport} 
Given a specific plasma we can in principle measure its equation of state as a function of the variables $T,\mu$. In addition, by evaluating the Kubo formulae \eqref{eq:Kubo} we can measure the various (bold) transport coefficients. We can then find a suitable frame by tuning the remaining parameters.

Holographic models provide a toy model for plasma physics. In particular \cite{Grozdanov:2017kyl} obtained the thermodynamics and transport properties for a specific plasma. In the weak field regime $T/\sqrt{\rho}\gg1$ the thermodynamics quantities take the form
\begin{equation}
    \label{eq:EoS_hol}
    \epsilon=a_0 T^4,~~p=a_1 T^4,~~ s=a_2 T^3,~~\mu=a_3 \rho~~,
\end{equation}
where the relation $a_2=a_0+a_1$ holds and the explicit values of the  remaining constants are
\begin{equation}
    a_0=\frac{N_c^2}{2\pi^2}~74.1,~~ a_1=\frac{N_c^2}{2\pi^2}~25.3,~~a_3=\frac{N_c^2}{2\pi^2} 10.9~~.
\end{equation}
Here $N_c$ is the number of colours and for clarity we set $N_c=1$. The expressions for the physical transport coefficients \eqref{eq:boldcoef} in turn read
\begin{equation} \label{eq:transport}
\begin{split}
    &\bs{\eta}_\parallel=\bs{\eta}_\perp=\frac{s}{4\pi}, ~~ \bs{r}_\parallel=\bs{r}_\perp=3.37~ \frac{\rho}{\mu T}~~,
    \\
    \bs{\zeta}_\perp=&\frac{1}{4}\bs{\zeta}_\parallel=0.33 ~\frac{s}{4\pi},~~\bs{\zeta}_\times=\bs{\zeta}'_\times=-\frac{1}{2}\bs{\zeta}_\parallel=-0.66~\frac{s}{4\pi}~~,
\end{split}
\end{equation}
which manifestly satisfy the entropy constraints \eqref{S_constr} and Onsager's relation \eqref{eq:CPT}.\footnote{The numerical factors appearing in \eqref{eq:EoS_hol} and \eqref{eq:transport} were obtained numerically in \cite{Grozdanov:2017kyl} and are approximate.}
In order to explicitly introduce the physical coefficients into our equations we decided to instead express them in terms of other parameters. In the weak field limit, these relations take the form
\begin{equation}
\begin{split}
    m_1&=\boldsymbol{r}_\perp+\mathcal{O}\left(\frac{\sqrt{\rho}}{T}\right)~~,
    \\
    \ell_1&=\boldsymbol{\eta}_\parallel+\mathcal{O}\left(\frac{\sqrt{\rho}}{T}\right)~~,
    \\
    f_1&=\boldsymbol{\zeta}_\perp-\frac{1}{3}\left[T\left(\frac{\varepsilon_3}{3}-f_3\right)-\varepsilon_1\right]+\mathcal{O}\left(\frac{\sqrt{\rho}}{T}\right)~~,
    \\
    f_2&=\boldsymbol{\zeta}_\times-\frac{1}{3}\left[T\left(\frac{\varepsilon_3}{3}-f_3\right)-\varepsilon_2\right]+\mathcal{O}\left(\frac{\sqrt{\rho}}{T}\right)~~,
    \\
    \tau_1&=\boldsymbol{\zeta}'_\times-\frac{1}{3}\left[T\left(\frac{\varepsilon_3}{3}-\tau_3\right)-\varepsilon_1\right]+\mathcal{O}\left(\frac{\sqrt{\rho}}{T}\right)~~,
    \\
    \tau_2&=\boldsymbol{\zeta}_\parallel-\frac{1}{3}\left[T\left(\frac{\varepsilon_3}{3}-\tau_3\right)-\varepsilon_2 \right]+\mathcal{O}\left(\frac{\sqrt{\rho}}{T}\right)~~.
\end{split}
\end{equation}
where we have already assumed the specific choice of frame \eqref{MS_choice1}. In what follows we found convenient
to rewrite equations in terms of dimensionless quantities. This can be done by taking care of the temperature scaling of the transport parameters and introducing the following dimensionless ratios
\begin{equation}
\label{eq: cnvrs}
    \begin{split}
        &G_1=\frac{1}{a_3 T n_2},~~G_2=\frac{(a_0+a_1)T^3}{k_2},~~G_3=\frac{(a_0+a_1)T^3}{\chi_1},~~G_4=3\frac{(a_0+a_1)T^2}{\varepsilon_3},~~G_5=\frac{1}{a_3\varrho_4},
        \\
        &X_1=a_3 T\boldsymbol{r}_\parallel,~~X_2=\frac{\boldsymbol{\eta}_\perp}{(a_0+a_1)T^3},~~X_3=\frac{\boldsymbol{\zeta}_\parallel}{(a_0+a_1)T^3},
        \\
        &Y_1=\frac{k_1}{(a_0+a_1)T},~~Y_2=a_3\frac{m_2}{T},~~Y_3=\frac{\ell_2}{(a_0+a_1)T},~~Y_4=a_3 \frac{n_1}{T},
        \\
        &Z_1=\frac{\varepsilon_2}{3(a_0+a_1)T^3},~~Z_2=\frac{\tau_3}{3(a_0+a_1)T^2},~~Z_3=a_3\frac{\varrho_2}{T},~~Z_4=\frac{\tau_4}{(a_0+a_1)T^2},
        \\
        &F_1=\frac{\varepsilon_1 }{(a_0+a_1)T^3},~~F_2=\frac{f_4}{(a_0+a_1)T^2},~~F_3=\frac{f_3}{(a_0+a_1)T^2},~~F_4=a_3\frac{\varrho_1}{T}~~.
    \end{split}
\end{equation}
These ratios together with the redefinitions 
\begin{equation}
    \delta=\frac{\Delta}{T}~~,~~\kappa=\frac{k}{T}~~,
\end{equation}
are sufficient to parametrise all terms in the polynomials. Notice that $X_i>0$ as a consequence of the entropy constraints \eqref{S_constr}. In addition, \eqref{eq:transport} implies that 
\begin{equation}
    X_1=3.37,~~X_2=\frac{1}{4\pi},~~X_3=\frac{1.33}{4\pi}.
\end{equation}
With these ratios we observe below that the quantities $G_i$ match the expressions for gapped modes both in the Alfvén and magnetosonic channels. The conditions we found in eq. \eqref{eq: MS_gaps_stability} imply that $G_{1,2,3}<0$ and $G_{4,5}<0$. 

As noted in \cite{Grozdanov:2017kyl}, this holographic plasma in the weak field limit tends to an isotropic conformal uncharged fluid but with signatures of anisotropy as the existence of a non-zero value of $r_\perp$ leads to anisotropic evolution. In addition, in a general frame, the coefficients $n_1$ and $n_2$ as well as $\chi_2$ (the latter two being necessary for the stability of gapped modes) break isotropy. This implies that even in the weak field limit we are faced with the non-trivial task of finding values for the transport coefficients such that stability and causality is attained for every angle $\theta$. We will now study constraints in both channels separately. 

\subsection{Alfvén channel}
\label{sec: Alfv_EoS}
In the weak field regime the spectral function in the Alfvén channel \eqref{Alfvén_poly} reduces to
\begin{equation}
    \left[\delta^2-G_1(\delta+X_1 \kappa^2)\right]\left[\delta^2-G_2(\delta+X_2\kappa^2)\right]+G_1 G_2 (Y_1+Y_3)(Y_2+Y_4)\delta^2 \kappa^2 \cos^2\theta~~.
\end{equation}
The two gapped modes simply become
\begin{equation}
    \omega=i T G_1,~~\omega=i T G_2+\mathcal{O}\left(\frac{\sqrt{\rho}}{T}\right)~~.
\end{equation}
Hence, a necessary condition for stability is that $G_{1,2}<0$. It is immediate to see that for $Y_1=-Y_3$ or $Y_2=-Y_4$ the polynomial factorizes. In particular, under these choices of the transport coefficients, one can deal away with the dependence on $\theta$. This enormously simplify the analysis, since one has to deal with just two quadratic polynomials, $P_{A_1}=\delta^2-G_1(\delta+X_1 \kappa^2)$ and $P_{A_2}=\delta^2-G_2(\delta+X_2 \kappa^2)$.
For either polynomials stability implies that
\begin{equation}
    \label{eq: Stability_AG}
    -G_i>0,~~ -G_i X_i >0 ~~ \text{for} ~~ i=1,2 ~~,
\end{equation}
which is always true as a consequence of the entropy constraints and gapped modes stability.
Causality instead amounts to impose
\begin{equation}
    \label{eq: Causality_AG}
    1>-G_i X_i>0 ~~ \text{for} ~~ i=1,2.
\end{equation}
One obvious possibility to satisfy this constraint is by choosing $G_i=-(q_i X_i)^{-1}$ for any real $q_i>1$. This concludes the analysis of stability and causality in the Alfvén channel.

\subsection{Magnetosonic channel}
In the weak field limit the gapped mode expressions \eqref{eq:gappedmag} are written in terms of $G_i$ defined in \eqref{eq: cnvrs} according to
\begin{equation}
\begin{split}
    \label{eq: Gapped_G}
    &\omega=i TG_1,~~\omega=iTG_2+\mathcal{O}\left(\frac{\sqrt{\rho}}{T}\right),~~\omega=iTG_3+\mathcal{O}\left(\frac{\sqrt{\rho}}{T}\right)~~,\\
    &\omega=iTG_4+\mathcal{O}\left(\frac{\sqrt{\rho}}{T}\right),~~\omega=iTG_5+\mathcal{O}\left(\frac{\sqrt{\rho}}{T}\right)~~.
    \end{split}
\end{equation}
We observe that \eqref{eq: cnvrs} implies that the gaps are stable under the conditions \eqref{eq:cc5}. We now consider the regime at arbitrary $k$. The analysis is complicated by the fact that the angle $\theta$ appears in the polynomial even in the weak field regime. We therefore present some analytic results for the longitudinal ($\theta=0$) and transverse ($\theta=\pi/2$) directions respectively but proceed and perform a numerical analysis at arbitrary angles.

\begin{figure}[h!]
  \centering
  \begin{minipage}[b]{0.45\textwidth}
    \includegraphics[width=\textwidth]{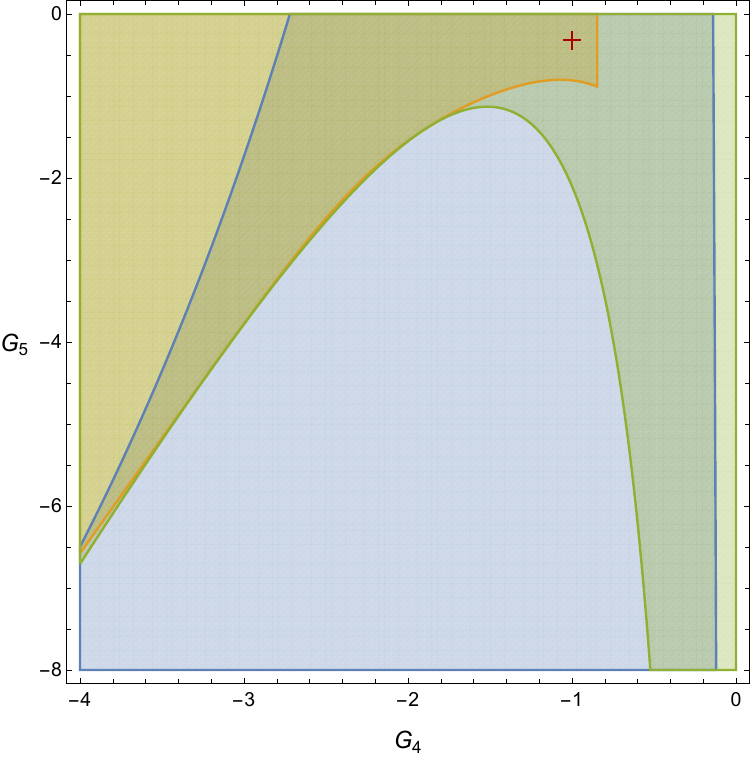}
  \end{minipage}
  \hfill
  \begin{minipage}[b]{0.45\textwidth}
    \includegraphics[width=\textwidth]{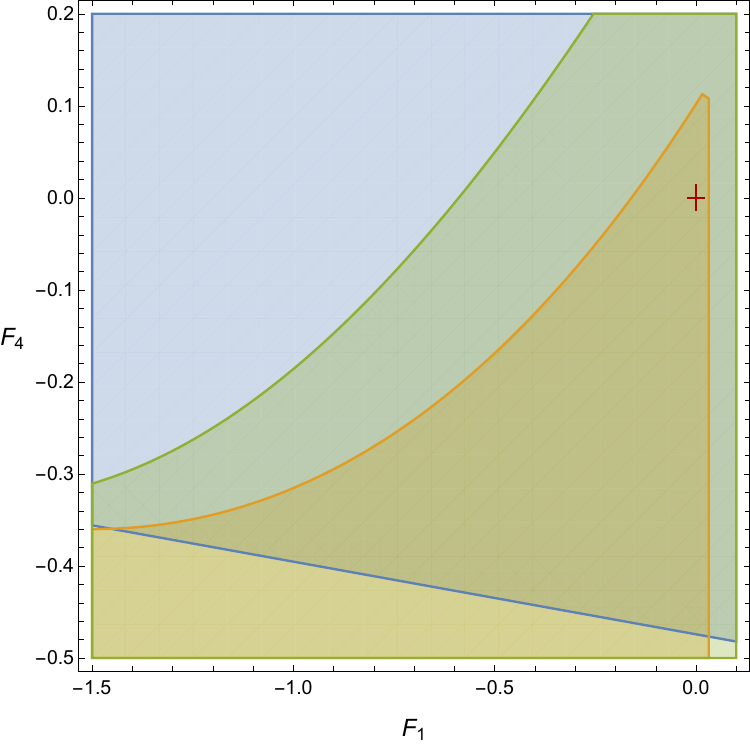}
  \end{minipage}
  \caption{
    The stability and causality subregions in the magnetosonic channel at fixed angles are depicted. We considered $\theta=0$ in the left panel and $\theta=\pi/2$ in the right panel. The plots are obtained by fixing $G_1=-\frac{1}{20 X_1}$, $G_2=-\frac{1}{20 X_2}$, $G_3=-\frac{1}{30 X_2}$, $Y_1=-Y_3=0$, $Y_2=1.35$, $Y_4=0$, $Z_1=0$, $Z_2=-0.5$, $Z_3=-1$, $Z_4=0.04$, $F_2=-0.35$ and $F_3=-1.35$. In the left panel we let $G_4$ and $G_5$ vary while in the right panel we vary $F_1$ and $F_4$.
    In both panels the orange represents the region where the stability conditions hold; the blue and green colors, respectively, represent causal regions. The superposition of the three colored regions implies that there exists a physical choice of hydrodynamics frame in which the dispersion relations are simultaneously stable and causal.
    \emph{Left panel:} The orange region is obtained using eq. \eqref{eq:cc5} and \eqref{eq: Stability_P5}; the green region via the first inequality in \eqref{eq: Causality_P5}; the blue region via the second and third condition in \eqref{eq: Causality_P5}. The plot shows the allowed parameter space by varying along $G_4$ and $G_5$. 
    The red cross marks the stable and causal point used in Fig. \ref{Fig:modes}, and corresponds to the values $G_4=-1$ and $G_5=-0.32$. \emph{Right panel}: the orange region is obtained via eq. \eqref{eq:cc5} and \eqref{eq: P6_Stability}; the green region via \eqref{eq: P6_real}; the blue region via \eqref{eq: P6_Causality}. The plot 
    shows the allowed parameter space by varying along $F_1$ and $F_4$.
    The red cross is the point chosen for the plots in Fig. \ref{Fig:modes}, corresponding to $F_1=F_4=0$.}
    \label{Fig:stab}
\end{figure}

In the longitudinal direction the polynomial \eqref{Poly_P9} in the weak field limit factorises into $P_9(\delta)=P_4(\delta)P_5(\delta)$ (see app.~\ref{app:furtherdetails}). In particular $P_4(\delta)$ is given by
\begin{equation}
    \label{eq: P4_coeffs_G}
    P_4(\delta)=\left[\delta^2-G_1(\delta+X_1 \kappa^2)\right]\left[\delta^2-G_2(\delta+X_2\kappa^2)\right]+G_1 G_2 (Y_1+Y_3)(Y_2+Y_4)\delta^2 \kappa^2~~,
\end{equation}
and matches the spectral function of the Alfvén channel in the weak field regime at $\theta=0$. All the considerations made in sec. \ref{sec: Alfv_EoS} for causality and stability apply here as well. The polynomial $P_5(\delta)$ is a bit more cumbersome and we provide the details of the coefficients in app.~\ref{app:furtherdetails}. In Fig.~\ref{Fig:stab} on the left we show that for a specific choice of coefficients there is a region of parameter space in which all constraints \eqref{eq:cc5}, \eqref{eq: Stability_P5}, \eqref{eq: Causality_P5} together with \eqref{eq: Stability_AG} and \eqref{eq: Causality_AG} are satisfied.

In the transverse direction the polynomial eq.~\eqref{Poly_P9} factories as $P_9(\delta)=P_6(\delta)P_3(\delta)$ (see app.~\ref{app:furtherdetails}) and $P_3(\delta)$ is given by
\begin{equation}
    \label{eq: P3_coeffs_G}
    P_3(\delta)=\delta^3-(G_1+G_3)\delta^2+G_3\left[G_1+\kappa^2(Y_3 Y_4 G_1-X_2)\right] \delta +\kappa^2 X_2 G_3 G_1~~.
\end{equation}
It is possible to find numerical values for the transport parameters such that the constraints given in app.~\ref{app:furtherdetails}, specifically \eqref{eq: P3_Stability} and \eqref{eq: P3_Causality} are satisfied. However, choosing a frame such that $Y_3=0$ the polynomial further factorizes into
\begin{equation}
    P_3(\delta)=(\delta-G_1)(\delta^2-G_3\delta-\kappa^2G_3 X_2)~~,
\end{equation}
and both polynomials are stable and causal with the following choices 
\begin{equation}
    G_1=-\frac{1}{q_1 X_I},~~ G_3=-\frac{1}{q_3 X_2} ~~\text{with} ~~ q_1,q_3>1~~.
\end{equation}
Notice that all these choices are consistent with stability and causality for both Alfvén modes and longitudinal magnetosonic modes discussed above. In turn, the polynomial $P_6(\delta)$ is more involved and we provide details in app.~\ref{app:furtherdetails}. In Fig.~\ref{Fig:stab} on the right panel we show a region of the parameter space where the stability and causality constraints, in particular eqs. \eqref{eq:cc5},\eqref{eq: P6_Stability},\eqref{eq: P6_real} and \eqref{eq: P6_Causality} are simultaneously satisfied. 

\begin{figure}[h!]
  \centering
  \begin{minipage}[b]{0.45\textwidth}
    \includegraphics[width=1\textwidth]{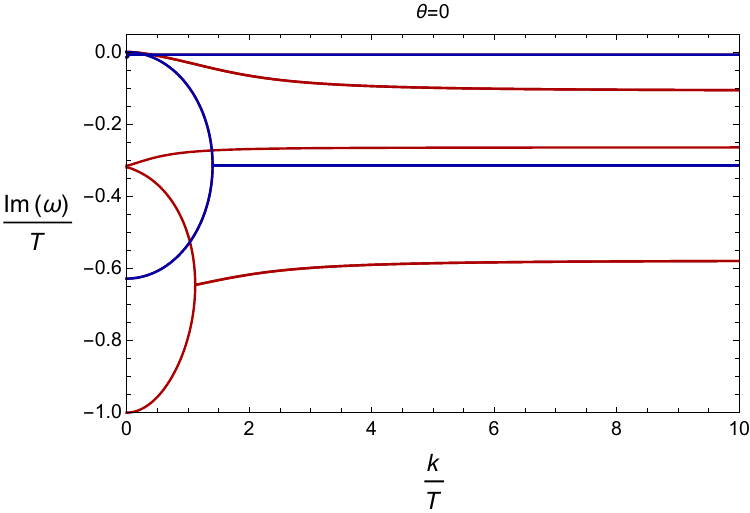}
  \end{minipage}
  \hfill
  \begin{minipage}[b]{0.45\textwidth}
    \includegraphics[width=1\textwidth]{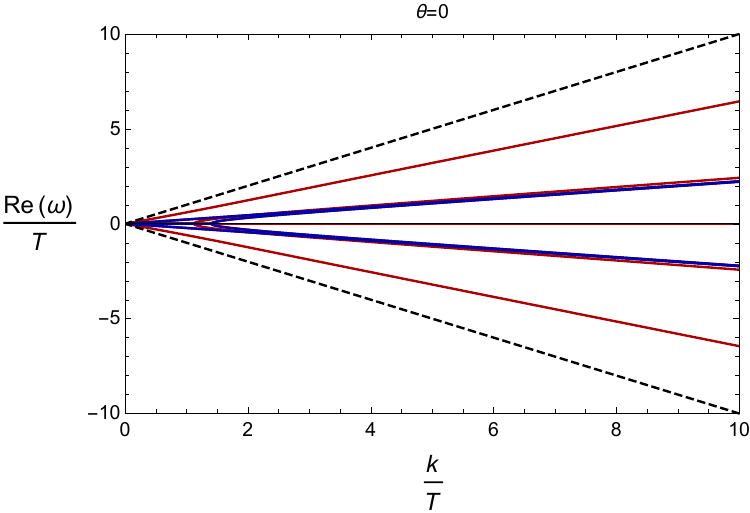}
  \end{minipage}
  \\
  \vspace{5mm}
  \begin{minipage}[b]{0.45\textwidth}
    \includegraphics[width=\textwidth]{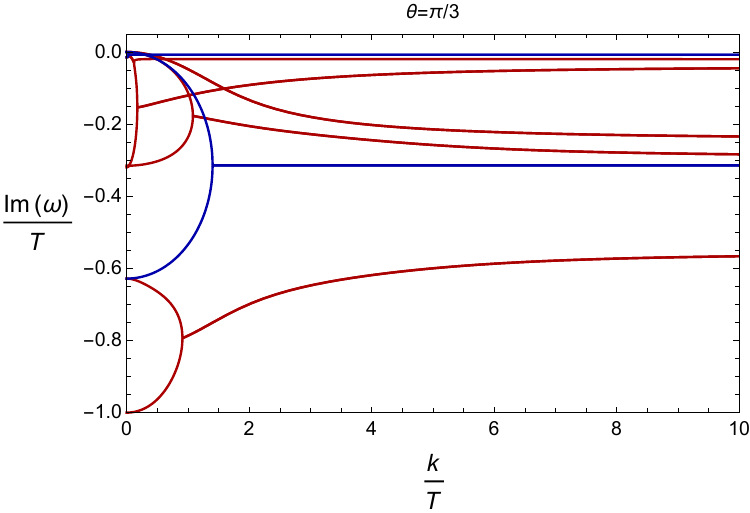}
  \end{minipage}
  \hfill
  \begin{minipage}[b]{0.45\textwidth}
    \includegraphics[width=\textwidth]{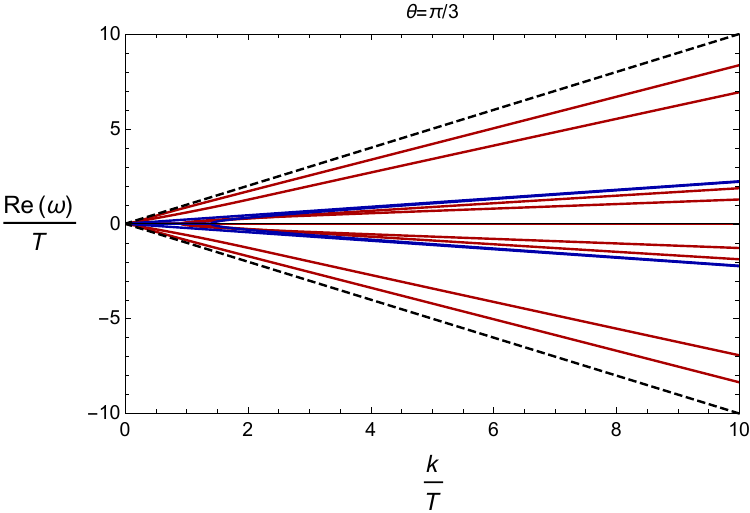}
  \end{minipage}
  \\
  \vspace{5mm}
  \begin{minipage}[b]{0.45\textwidth}
    \includegraphics[width=\textwidth]{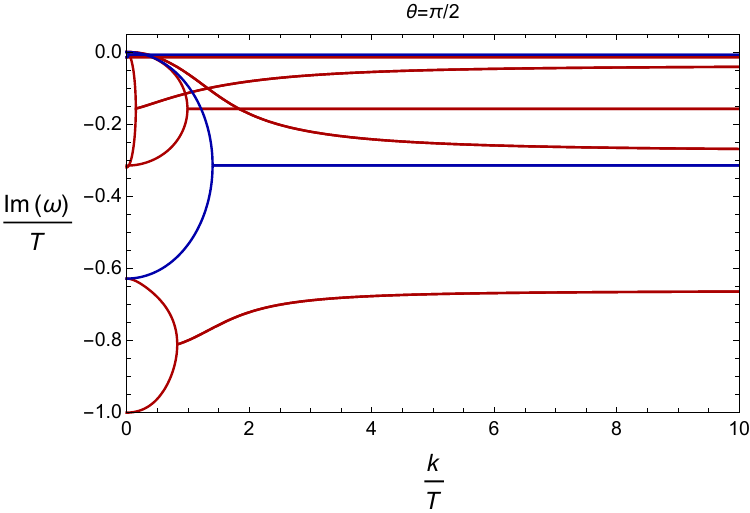}
  \end{minipage}
  \hfill
  \begin{minipage}[b]{0.45\textwidth}
    \includegraphics[width=\textwidth]{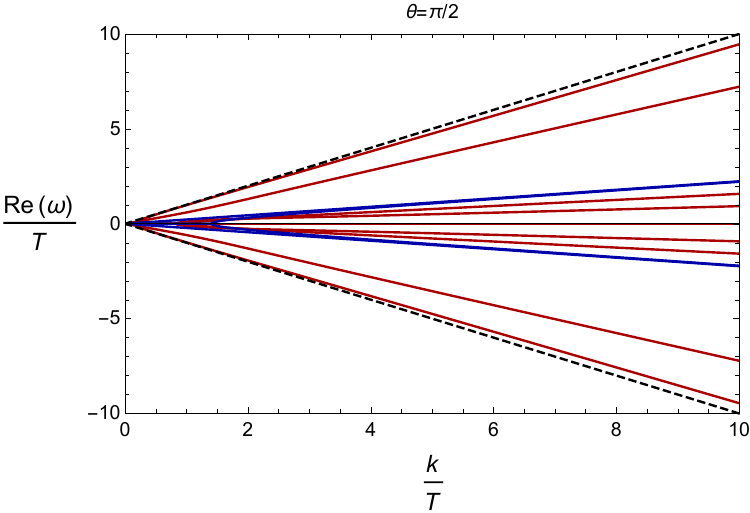}
  \end{minipage}
  \caption{
    Here we plot the real and imaginary parts of $\omega(k)$ for the angles $\theta=0,\pi/3,\pi/2$. The Alfvén modes are plotted in blue and the magnetosonic modes in red. The values for the parameters are: $G_1=-\frac{1}{20 X_1}$, $G_2=-\frac{1}{20 X_2}$, $G_3=-\frac{1}{30 X_2}$, $G_4=-1$, $G_5=-0.32$,  $Y_1=0$, $Y_2=1.35$, $Y_3=0$, $Y_4=0$, $Z_1=0$, $Z_2=-0.5$, $Z_3=-1$, $Z_4=0.04$, $F_1=0$, $F_2=-0.35$, $F_3=-1.35$, $F_4=0$. \emph{Left panel}: The imaginary part of $\omega(k)$ is plotted. The gapped modes, eq. \eqref{eq: Gapped_G} , are clearly visible and all curves are in the lower half-plane, meaning that the modes are stable. \emph{Right panel}: The real part of $\omega(k)$ is plotted. The dashed curves represent the light cone $\pm k$, and it is immediate to see that all curves are contained in the light cone and so respect causality. Notice also that all the curves are linear for $k\gg1$.}
    \label{Fig:modes}
\end{figure}

At arbitrary angles the polynomial \eqref{Poly_P9} does not in general factorise and so we perform a numerical analysis by scanning parameter space such that the positive imaginary part of all modes is minimised and the distance $(1-\lim_{k\to\infty}\left| \frac{\text{Re}\left(\omega(k)\right)}{k}\right|)$ is maximised. Indeed we are able to find a large region of parameter space in which this is the case. Within that region it is possible to satisfy stability and causality constraints at all angles even when setting $Z_1,Y_1,Y_3,Y_4,F_1,F_4$ to zero. In turn this implies, via eq.~\eqref{eq: cnvrs}, that there is a choice of frame that satisfies all constraints for which
\begin{equation} \label{eq:newc}
k_1=0~~,~~\ell_2=0~~,~~n_1=0~~,~~\varepsilon_1=0~~,~~\varepsilon_2=0~~,~~\varrho_1=0~~.  
\end{equation}
Together with the choice \eqref{MS_choice1}, this means that besides the 7 coefficients characterising transport \eqref{eq:transport} in which CPT invariance was imposed, 11 additional coefficients are need to ensure causal and stable evolution. In Fig. \ref{Fig:modes} we give examples of dispersion relations for various angles in the case \eqref{eq:newc}. It is clear from Fig. \ref{Fig:modes} that all modes are stable and causal. The domain of causality and stability that we found allows to vary some of the parameters, for instance
\begin{equation}
\begin{split}
&0\le Y_1\lesssim 0.05~~,~~0\le Y_4\lesssim 0.02~~,~~0\le F_1\lesssim 0.02~~,\\
&0\le F_4\lesssim 0.08~~,~~0\le Z_1\lesssim 0.03~~,~~0.04\le Z_4\lesssim 0.1~~,\\
&1.32\lesssim F_3 \lesssim 1.4~~,~~0\lesssim Y_3\lesssim0.1~~,~~0.32\lesssim G_5\lesssim 0.4~~,
\end{split}
\end{equation}
such that causality and stability for all $\theta$ is still preserved. Therefore the specific point of parameter space depicted in Fig. \ref{Fig:modes} is not exceptional and a much larger family of suitable frames was identified. The dimensionless ratios \eqref{eq: cnvrs} determine the dependence of the various transport coefficients with the temperature, for instance 
\begin{equation}
\tau_4=\frac{4}{100}(a_0+a_1)T^2~~,    
\end{equation}
for the specific choice in Fig.~\ref{Fig:modes} and similarly for the remaining transport coefficients.

It is possible that the parameter set that we found could be reduced further as we did not perform an extensive search of all possible parameter excursions. On the other hand, since the theory in the Landau frame already involves 7 transport coefficients, we are not surprised that an additional 11 coefficients are needed. In order to place this number into perspective, we point out that Ref.~\cite{Hoult:2020eho} noted that a conformal isotropic charged fluid with only 2 parameters in the Landau frame requires more than the double (at least 3) additional parameters to ensure stability and causality.

\section{Discussion} \label{sec:discussion}
In this work we formulated MHD in a general frame at first order in gradients, under the assumptions of parity-invariance and discrete charge symmetry, and determined general constraints on transport coefficients that render the theory linearly stable and causal for generic equilibrium states with constant magnetic fields. As these constraints are unwieldy analytically due to the sheer number of transport coefficients involved, our strategy consisted in providing a method by which for a given plasma (with equation of state and transport properties given by Kubo formulae \eqref{eq:Kubo}) one can scan parameter space numerically for suitable frames. Once a choice of stable and causal frame is identified, the model is fixed and one can proceed with analysing it in different contexts (e.g. by choosing different types of initial data and study their evolution). 

We applied the method outlined above to a specific equation of state arising from a holographic toy model \cite{Grozdanov:2017kyl} in sec.~\ref{sec:EOS} and found that besides the 7 transport coefficients obtained via Kubo formulae, an additional 11 parameters are needed to ensure stable and causal linear perturbations at every angle $\theta$.\footnote{Our analysis of the equation of state of \cite{Grozdanov:2017kyl}  was performed in the weak field limit in which signatures of anisotropy are small but significant. It would be interesting to also explore the strong field limit in which features of anisotropy are more salient.} We have not explored the region of parameter space in full generality and as such it is possible that causality and stability could be enforced with a reduced number of additional coefficients. Regardless, our work shows that applying the BDNK method \cite{Bemfica:2017wps, Bemfica:2019knx, Kovtun:2019, Hoult:2020eho, Bemfica:2020zjp} to MHD can be accomplished successfully.\footnote{There is a difference between our analysis and earlier applications of the BDNK method. In particular, the MHD case is similar to that of a superfluid in which equilibrium states are degenerate due to the last equation in \eqref{eq:conditions}. We have focused in the case of constant magnetic fields but other spatial dependent configurations are possible. In full generality the orientation of the magnetic field can be written as $h^\mu=(0,0,0,1)+\mathcal{O}(\partial)$. The analysis presented here ignores the contribution $\mathcal{O}(\partial)$. The constraints we obtained are nevertheless required but it is possible that when including the correction $\mathcal{O}(\partial)$ these constraints become stricter.} The model presented here can be used for studying various processes of interest in astrophysics and heavy-ion collisions and presents an interesting alternative to second order MHD theories such as \cite{Most:2021uck}.

Developing a systematic numerical method for scanning parameter space and identifying suitable frames would be useful, in particular when addressing more complicated scenarios.\footnote{Such method would be akin to those employed in the context of the conformal bootstrap programme (see e.g. \cite{Poland:2018epd}).} For instance, our analysis assumed parity-invariance and discrete charge symmetry but it would be interesting to consider situations in which these assumptions are relaxed and other effects are accounted for such as additional baryon currents, accretion via radiation fluxes, ambipolar diffusion and Hall drift \cite{Abramowicz:2011xu, ambipolar2021}. This generalisation would be timely given the relevance of the chiral magnetic effect both in heavy-ion collisions \cite{Zhao:2019hta} and astrophysics \cite{Rogachevskii:2017uyc}. However, such scenarios require working with many additional transport coefficients compared with the number we considered in this work (see e.g. \cite{Armas:2018zbe}) and a systematic numerical analysis would be necessary. We plan on addressing this in a future publication.

A particular case that we did not address here is the $T/\sqrt{\rho}\to0$ limit which is related to force-free electrodynamics (FFE) \cite{Grozdanov:2016tdf, Gralla:2018kif, Glorioso:2018kcp, Benenowski:2019ule, Poovuttikul:2021fdi}. As a first step, we would consider the strong magnetic field limit of the equation of state that we studied in sec.~\ref{sec:EOS}. It would be interesting to study stability of FFE configurations using the same methods employed here.  

It would also be of value to prove stability and causality at the nonlinear level and couple it to dynamical gravity as in \cite{Bemfica:2017wps, Bemfica:2020zjp, Hoult:2020eho}. We have not performed this study but previous cases suggest that the same constraints found at the linear level are sufficient for enforcing causality and stability at the nonlinear level. 

An interesting open question is to understand whether choices of frame (and the additional transport coefficients necessary for causality and stability) in the context of MHD can be derived from some microscopic theory, perhaps along the lines of \cite{Bemfica:2017wps, Hoult:2021gnb}. Kinetic theory has already been used to derive other formulations of MHD \cite{Denicol:2018rbw}.

As a concluding note, we envision that this study could give rise to many applications, including understanding the mode spectra of rotating magnetised stars \cite{Armas:2018atq}, numerical studies comparing first and second order models \cite{Pandya:2021ief} as well as numerical simulations of the large scale structure of spacetime, and their imprint into primordial gravitational waves \cite{Brandenburg:2021aln}.

\section*{Acknowledgements}
We would like to thank A. Cole, A. Jain, P. Kovtun, J. Noronha and N. Poovuttikul for useful discussions. JA is partly
supported by the Netherlands Organization for Scientific
Research (NWO) through the NWA Startimpuls funding scheme and by the Dutch Institute for Emergent Phenomena (DIEP) cluster at the University of Amsterdam. JA acknowledges the Niels Bohr Institute for hospitality during several stages of this project.

\appendix
\section{Details on one-form hydrodynamics} \label{app:details1form}
In this appendix we give further details on one-form hydrodynamics following \cite{Armas:2018zbe}. The comments and clarifications made here are not necessary for understanding the core of the paper. We begin with a discussion of the equilibrium partition function and then summarise the transformations of the relevant MHD fields under various discrete symmetries.

\subsection{Equilibrium partition function}
We consider one-form hydrodynamics coupled to a background metric $g_{\mu\nu}$ and a two-form gauge field $b_{\mu\nu}$. These background fields transform under infinitesimal diffeomorphisms parametrised by $\chi^\mu$ and one-form gauge transformations parametrised by $\Lambda^\chi_\mu$ according to
\begin{equation}
\delta_\mathfrak{X} g_{\mu\nu}=\mathcal{L}_{\chi}g_{\mu\nu}~,~~\delta_\mathfrak{X} b_{\mu\nu}=\mathcal{L}_{\chi}b_{\mu\nu}+2\partial_{[\mu}\Lambda^\chi_{\nu]}~~,
\end{equation}
where $\mathfrak{X}=(\chi^\mu,\Lambda^\chi_\mu)$ and where $\mathcal{L}_\chi$ denotes the Lie derivative along $\chi^\mu$. In turn, the hydrodynamic fields are described by the set $B=(\beta^\mu,\Lambda^{\beta}_\mu)$. These fields can be expressed in terms of the more usual hydrodynamic fields, in particular fluid velocity $u^\mu$, temperature $T$ and one-form chemical potential $\mu_\mu$ according to
\begin{equation}
\beta^\mu=\frac{u^\mu}{T}~~,~~\Lambda^{\beta}_\mu+\beta^\nu b_{\nu\mu}=\frac{\mu_\mu}{T}~~.
\end{equation}
However, it is straightforward to see that $\mu_\mu$ is not invariant under one-form gauge transformations. In order to define a gauge-invariant one-form chemical potential it is required to include another dynamical field, in particular a scalar "Goldstone" $\varphi$ akin to a magnetic scalar potential that transforms according to
\begin{equation}
 \delta_\mathfrak{X}\varphi=\mathcal{L}_{\chi}\varphi-\beta^\mu\Lambda_\mu^\chi~~.
\end{equation}
This is equivalent to state that the one-form symmetry is spontaneously broken along the fluid velocity $u^\mu$ \cite{Armas:2018atq,Armas:2018zbe}.
Given this transformation, we can define a gauge-invariant vector such that
\begin{equation}
\mu h_\mu=\mu_\mu-T\partial_\mu\varphi~~,    
\end{equation}
where $h^\mu$ is normalised such that $h^\mu h_\mu=1$. We note that the one-form chemical potential $\mu_\mu$ encodes both the chemical potential $\mu$ and the vector $h^\mu$. It may be observed that $h^\mu u_\mu=-T^2\delta_B \varphi/\mu$ and hence $u^\mu$ and $h_\mu$ are not orthogonal at this point. However, it may be shown, using the adiabaticity equation (see \cite{Armas:2018zbe}) that $\delta_B \varphi=\mathcal{O}(\partial)$. One is free to use the redefinition freedom associated with $\mu_\mu$, namely $\mu_\mu\to\mu_\mu +\delta\mu_\mu$ to set $\delta_B \varphi=0$ at all orders, which is a type of Josephson condition. This means that we are free to take $h^\mu u_\mu=0$ as an exact all order statement. This is the choice we made throughout this paper, which leads to the physical interpretation of the fluid field variables $\mu$ and $h^\mu$ being related to the strength and the direction of the magnetic field lines respectively. However we note that it is possible not to use this redefinition freedom and to consider an even larger family of MHD frames. 

Given these considerations we now discuss aspects of equilibrium partition functions in one-form hydrodynamics. The on-shell equilibrium partition function, that is the equilibrium partition function in the grand canonical ensemble for which $\delta_B \varphi=0$ and $\varphi$ obeys the constraint $\nabla_\mu\left(T\rho h^\mu\right)=0$, can be written as 
\begin{equation} \label{eq:eqg}
\mathcal{Z}[g_{\mu\nu},b_{\mu\nu}]=\text{exp}\int_\Sigma d\sigma_\mu\left(T^{\mu\nu}K_\nu +(\Lambda^K_\nu+K^\lambda b_{\lambda\nu})J^{\mu\nu}\right)~~,   
\end{equation}
where $K^\mu$ is a Killing vector field and $\Lambda^K_\mu$ is such that $\delta_B g_{\mu\nu}=\delta_B b_{\mu\nu}=0$. In addition, $\Sigma$ denotes an arbitrary Cauchy slice with volume element $d\sigma_\mu$. For the specific equilibrium states that we consider in Sec.~\ref{sec:equilibrium} with $g_{\mu\nu}=\eta_{\mu\nu}$ and $b_{\mu\nu}=0$ we have that $K^{\mu}=\delta^\mu_t/T_0$ and $\Lambda_\mu^K=\mu_0/T_0 \delta_\mu^z$ where $T_0$ and $\mu_0$ denote the constant global temperature and chemical potential respectively. In the case of these global thermal states, and given that all first order corrections \eqref{eq:param} vanish in equilibrium, the grand canonical partition function becomes
\begin{equation} \label{eq:eq}
    \mathcal{Z}[g_{\mu\nu}=\eta_{\mu\nu},b_{\mu\nu}=0]=\text{exp}\int_\Sigma d\sigma_\mu u^\mu\frac{1}{T_0}\left(-\varepsilon_0+\mu_0\rho_0\right)~~,  
\end{equation}
which is the same as encountered in the context of zero-form hydrodynamics with energy $\varepsilon_0$, chemical potential $\mu_0$ and “particle number" $\rho_0$. We note that $\mu_0$ is the value of the z-component of the one-form chemical potential $\mu_\mu$ for equilibrium configurations with $h^\mu=(0,0,0,1)$, which are the ones considered in this paper. Given the partition function, we require the variances for energy and “particle number" fluctuations to be positive for the equilibrium thermal state, in particular $\Delta \varepsilon^2=\langle \varepsilon^2\rangle-\langle \varepsilon\rangle^2>0$, $\Delta \rho^2>0$ and $\Delta(\varepsilon \rho)^2>0$. These conditions lead to the last three conditions in \eqref{eq:assthermal}.

In addition we note that the partition function \eqref{eq:eq} is invariant under discrete one-form charge symmetry, according to the transformations that we provide in subsection \ref{app:discrete} below. This is due to the fact it it only depends on $T$ and $\mu$, the latter being determined in terms of the modulus of the magnetic fields. This is not the case generically at higher orders in the gradient expansion. Under the assumption that the underlying microscopic field theories are well behaved we can compute \eqref{eq:eqg} using an Euclidean path integral of the form
\begin{equation}
\mathcal{Z}[g_{\mu\nu},b_{\mu\nu}]=\int \mathcal{D}\varphi\exp\left(-S^{\text{HS}}[g_{\mu\nu},b_{\mu\nu};\varphi]\right)~~,
\end{equation}
where $S^{\text{HS}}$ is the equilibrium (hydrostatic) effective action, which can be parametrised in terms of a free energy current $N^\mu_{\text{HS}}$ such that
\begin{equation}
S^{\text{HS}}[g_{\mu\nu},b_{\mu\nu};\varphi]=\int_{\Sigma}d\sigma_\mu N^\mu_{\text{HS}}~~.
\end{equation}
For one-form fluids describing MHD, the component of the free energy current along $u^\mu$ takes the form \cite{Armas:2018zbe}
\begin{equation} \label{eq:ncurrent}
\begin{split}
u_\mu N^\mu_{\text{HS}}=-\frac{1}{T}\big(&p-\frac{\alpha}{6}e^{\mu\nu\rho\sigma}u_\mu H_{\nu\rho\sigma}-\beta e^{\mu\nu\rho\sigma}u_\mu h_\nu \partial_\rho u_\sigma-\tilde \beta_1 h^\mu\partial_\mu T\\
&-\tilde \beta_2 h^\mu\partial_\mu\frac{\mu}{T}-\tilde \beta_3 e^{\mu\nu\rho\sigma} u_\mu h_\nu\partial_\rho h_\sigma\big) +\mathcal{O}\left(\partial^2\right)~~,
\end{split}
\end{equation}
where $p(T,\mu)$ is the pressure and $\alpha, \beta, \tilde \beta_1, \tilde \beta_2, \tilde \beta_3$ are transport coefficients, all functions of $T,\mu$.\footnote{We have only given the components of the free energy current along $u^\mu$ in Eq.~\eqref{eq:ncurrent} for clarity of presentation. The components orthogonal to it also vanish when discrete one-form charge and parity symmetries are imposed \cite{Armas:2018zbe}. Note that $\mu$ in this work corresponds to $\varpi$ in \cite{Armas:2018zbe}.} In particular, $\alpha=\nu_0$ (see Eq. (7.48) of \cite{Armas:2018zbe}) with $\nu_0$ being the $\mathcal{O}(1)$ term in the expansion of the electric chemical potential $\nu=\nu_0+\mathcal{O}(\partial)$. Given the transformation properties of the various fields under discrete one-form charge symmetry C and parity P, enforcing such symmetries leads to $(\alpha, \beta, \tilde \beta_1, \tilde \beta_2, \tilde \beta_3)=0$ and as a consequence $\nu=\mathcal{O}(\partial)$. In turn, requiring the same symmetries implies that $q(T,\nu)\sim\mathcal{O}(\partial^2)$ (see Eq. (7.23) and (7.48) of \cite{Armas:2018zbe}) and hence in the bulk of the paper we have restricted to fluid configurations with vanishing electric charge density up to $\mathcal{O}(\partial^2)$.

\subsection{Transformations under discrete symmetries} \label{app:discrete}
In this section we provide details about the transformations of the various fields entering the formulation of magnetohydrodynamics for the discrete symmetries of charge conjugation $\text{C}$, parity $\text{P}$ and time-reversal $\text{T}$. Table \ref{tab:CPT} summarises the transformation properties of the various fields. The interpretation of discrete symmetries in this one-form formulation of MHD is in general not the same as in traditional formulations of MHD. As explained in appendix C of \cite{Armas:2018zbe}, denoting $\text{C}_{\text{EM}}$, $\text{P}_{\text{EM}}$ and $\text{T}_{\text{EM}}$ for charge-conjugation, parity and time-reversal operations in the traditional formulation we obtain the following map
\begin{equation}
\text{C}_{\text{EM}}=\text{C}~~,~~\text{P}_{\text{EM}}=\text{C}\text{P}~~,~~\text{T}_{\text{EM}}=\text{C}\text{T}~~.    
\end{equation}
In particular we note that the operation of one-form charge conjugation is equivalent to the operation of electric charge conjugation in the traditional formulation for MHD fields. This is due to the fact that magnetic fields flip sign under electric charge conjugation as well as under one-form charge conjugation. Throughout this work, we assumed that the theory is invariant under $\text{C}$ and $\text{P}$ individually, which captures the sector of parity-invariant MHD with sub-leading electric chemical potential $\nu$. 

\begin{table}[t]
  \begin{subtable}[t]{0.31\textwidth}
    \flushright
    \begin{tabular}[t]{ccccc}
    \\
      \toprule
      & C & P & T & CPT \\
      \midrule
      $T^{tt}$, $g_{tt}$ & $+$ & $+$ & $+$ & $+$ \\
      $T^{ti}$, $g_{ti}$ & $+$ & $-$ & $-$ & $+$ \\
      $T^{ij}$, $g_{ij}$ & $+$ & $+$ & $+$ & $+$ \\
      \midrule
      $J^t$, $A_t$      & $-$ & $+$ & $+$ & $-$ \\
      $J^i$, $A_i$      & $-$ & $-$ & $-$ & $-$ \\
      \midrule
      $J^{ti}$, $b_{ti}$ & $-$ & $-$ & $+$ & $+$ \\
      $J^{ij}$, $b_{ij}$ & $-$ & $+$ & $-$ & $+$ \\
      \bottomrule
    \end{tabular}
  \end{subtable}
  \begin{subtable}[t]{0.3\textwidth}
    \centering 
    \begin{tabular}[t]{ccccc}
    \\\\
      \toprule
      & C & P & T & CPT \\
      \midrule
      $u^t$ & $+$ & $+$ & $+$ & $+$ \\
      $u^i$ & $+$ & $-$ & $-$ & $+$ \\
      $T$   & $+$ & $+$ & $+$ & $+$ \\
      \midrule
      $\nu$ & $-$ & $+$ & $+$ & $-$ \\
      $\mu_t$ & $-$ & $+$ & $-$ & $+$ \\ 
      $\mu_i$   & $-$ & $-$ & $+$ & $+$ \\
      $\varphi$ & $-$ & $+$ & $+$ & $-$ \\
      \bottomrule
    \end{tabular}
  \end{subtable}
  \begin{subtable}[t]{0.36\textwidth}
    \flushleft
    \begin{tabular}[t]{ccccc}
    \\ \\
      \toprule
      & C & P & T & CPT \\
      \midrule
      $F_{ti}$, $E_i$, $B^t$  & $-$ & $-$ & $+$ & $+$ \\
      $F_{ij}$, $B^i$, $E^t$ & $-$ & $+$ & $-$ & $+$ \\
      \midrule
      $\mu$ & $+$ & $+$ & $+$ & $+$ \\
      $h_{i}$  & $-$ & $-$ & $+$ & $+$\\
      $h_t$ & $-$ & $+$ & $-$ & $+$\\
      $H_{tij}$ & $-$ & $+$ & $+$ & $-$ \\
      $H_{ijk}$ & $-$ & $-$ & $-$ & $-$ \\
      \bottomrule
    \end{tabular}
  \end{subtable}
  \caption{Transformation properties of various quantities under the discrete
    symmetries C, P, and T. The first table summarises properties of conserved
    currents and the associated sources, the second table of dynamical fields,
    while the third table of various derived quantities. This table was adapted from \cite{Armas:2018zbe}.}
  \label{tab:CPT}
\end{table}

\section{Landau frame}\label{app:Landauframe}
The Landau frame is defined by setting $\varepsilon_i=\chi_i=k_i=\varrho_i=n_i=0$ or by finding a frame transformation that sets $\delta\varepsilon=\delta\chi=k^\mu=\delta\varrho=n^\mu=0$. In order to do so, we make use of \eqref{eq:framchange}, which requires
\begin{equation}
\begin{split}
&\tilde\gamma_i=-\frac{\chi_i}{Ts}~~,~~\theta_i=-\frac{k_i}{(\epsilon+P)}~~,~~\gamma_i=\frac{n_i}{\rho}~~,~~\omega_i=\frac{\varrho_i\frac{\partial\epsilon}{\partial T}-\varepsilon_i \frac{\partial\rho}{\partial T}}{\frac{\partial\rho}{\partial T}\frac{\partial\epsilon}{\partial\mu}-\frac{\partial\rho}{\partial \mu}\frac{\partial\epsilon}{\partial T}}~~,\\
&t_i=-\left(\frac{\partial\epsilon}{\partial T}\right)^{-1}\left(\frac{\partial\epsilon}{\partial\mu}\omega_i+\varepsilon_i\right)~~.
 \end{split} 
\end{equation}
The usual Landau frame, as found in \cite{Armas:2018atq, Armas:2018zbe}, also requires imposing the equations of motion \eqref{eq:eomdual}. Once they are imposed, the non-zero components of the stress tensor and charge current \eqref{eq:stress1} are given by
\begin{equation} \label{eq:paramLandau}
\begin{split}
&\delta f=-\boldsymbol{\zeta}_\perp\Delta^{\mu\nu}\nabla_\mu u_\nu -\boldsymbol{\zeta}_\times h^\mu h^\nu\nabla_\mu u_\nu~~,\\
&\delta \tau=-\boldsymbol{\zeta}_\times^{'}\Delta^{\mu\nu}\nabla_\mu u_\nu -\boldsymbol{\zeta}_{||} h^\mu h^\nu\nabla_\mu u_\nu~~,\\
&\ell^\mu=-T\boldsymbol{\eta}_{||}\Delta^{\mu\nu}\Delta^{\mu\sigma}h^{\nu}\delta_B g_{\nu\sigma}~~,\\
&m^\mu=-T\boldsymbol{r}_{\perp}\Delta^{\mu\nu}h^\lambda\delta_B b_{\nu\lambda}~~, \\
&t^{\mu\nu}=-T\boldsymbol{\eta}_{\perp}\left(\Delta^{\mu\rho}\Delta^{\nu\sigma}-\frac{1}{2}\Delta^{\mu\nu}\Delta^{\rho\sigma}\right)\delta_Bg_{\rho\sigma}~~,\\
&s^{\mu\nu}=-T \boldsymbol{r}_{||}\Delta^{\mu\rho}\Delta^{\nu\sigma}\delta_B b_{\rho\sigma}~~,
\end{split}
\end{equation}
where the transport coefficients in bold were defined in \eqref{eq:boldcoef}.

\section{Equilibrium states with finite spatial velocity} \label{app:boostedframe}
In the bulk of the paper we also discuss states with finite constant velocity. Such states can be obtained by performing a Lorentz boost, with boost velocity $\boldsymbol{\beta}$, starting from equilibrium configurations with zero velocity, as introduced in Sec.~\ref{sec:equilibrium}. Similarly, the dispersion relations found throughout Sec.~\ref{sec:constraints} for configurations with zero velocity can be "boosted" to dispersion relations characterising states with finite velocity. To wit, the fluid velocity for a state with vanishing spatial velocity is related to the fluid velocity of the state with non-vanishing spatial velocity by a Lorentz transformation
\begin{equation}
    u^\mu={\Lambda^\mu}_\nu u'^\nu~~\text{ with }~~u'^\mu=\gamma(1,\boldsymbol{\beta})~~,~~\gamma=\frac{1}{\sqrt{1-\bs\beta^2}}~~,
\end{equation}
where ${\Lambda^\mu}_\nu$ is a matrix in SO(1,3), $\bs\beta^2$ is the modulus of the boost velocity squared and $u'^\mu$ is the fluid velocity for the state with non-vanishing spatial velocity. Analogously, we introduce the wave vector $k'^\mu$ and direction of magnetic field lines $h'^\mu$ in the state with non-vanishing spatial velocity such that
\begin{equation}
    k'^\mu=(\omega',\boldsymbol{k}')~~,~~ h'^\mu=(h_t',\boldsymbol{h}')~~.
\end{equation}
These quantities are related to the respective quantities in the frame with vanishing velocity $(\omega,\boldsymbol{k}, h_t, \boldsymbol{h})$ according to the following relations
\begin{align}
    \label{boost_kh}
    &\begin{cases}
        &\omega=\gamma(\omega'-\boldsymbol{\beta}\cdot \boldsymbol{k}')\\
        &\boldsymbol{k}=\boldsymbol{k}'+\gamma\left(\frac{\gamma}{1+\gamma}\boldsymbol{\beta}\cdot\boldsymbol{k}'-\omega'\right)\boldsymbol{\beta}
    \end{cases}~~~~,
    &\begin{cases}
        &h_t=0=\gamma(h_t'-\boldsymbol{\beta}\cdot \boldsymbol{h}')\\
        &\boldsymbol{h}=\boldsymbol{h}'+\gamma\left(\frac{\gamma}{1+\gamma}\boldsymbol{\beta}\cdot\boldsymbol{h}'-h_t'\right)\boldsymbol{\beta}
    \end{cases}~~.
\end{align}
The original rest frame configuration satisfies $u^\mu h_\mu=0$. To ensure that this is the case in the new (boosted) frame it is sufficient to require that $\boldsymbol{\beta}\cdot\boldsymbol{h'}=0$. Eq.~\eqref{boost_kh} then implies that $h_t'=0$ and $\boldsymbol{h'}=\boldsymbol{h}$. As a consequence, for this class of frames one has that
\begin{equation}
    \boldsymbol{k}\cdot\boldsymbol{h}\equiv k\cos\theta=\boldsymbol{k}'\cdot\boldsymbol{h}=\boldsymbol{k}'\cdot\boldsymbol{h}'~~,
\end{equation}
where $k$ is given as a function of $k'$ and $\omega'$ according to
\begin{equation}
    \label{k2}
    k^2=\boldsymbol{k}\cdot\boldsymbol{k}=k'^2+2\gamma\left(\frac{\gamma}{1+\gamma}\boldsymbol{\beta}\cdot\boldsymbol{k}'-\omega'\right)\boldsymbol{\beta}\cdot\boldsymbol{k}'+\gamma^2\left(\frac{\gamma}{1+\gamma}\boldsymbol{\beta}\cdot\boldsymbol{k}'-\omega'\right)^2\bs\beta^2~~.
\end{equation}
Therefore, in order to obtain dispersion relations for states with non-vanishing constant spatial velocity, it is sufficient to make the following substitutions
\begin{equation}
    \label{sub}
    \begin{aligned}
        &\omega\longrightarrow\gamma(\omega'-\boldsymbol{\beta}\cdot \boldsymbol{k}')~~,\\
        &\boldsymbol{k}\longrightarrow\boldsymbol{k}'+\gamma\left(\frac{\gamma}{1+\gamma}\boldsymbol{\beta}\cdot\boldsymbol{k}'-\omega'\right)\boldsymbol{\beta}~~,\\
        &k^2\cos^2\theta\longrightarrow \left[k'^2+2\gamma\left(\frac{\gamma}{1+\gamma}\boldsymbol{\beta}\cdot\boldsymbol{k}'-\omega'\right)\boldsymbol{\beta}\cdot\boldsymbol{k}'+\gamma^2\left(\frac{\gamma}{1+\gamma}\boldsymbol{\beta}\cdot\boldsymbol{k}'-\omega'\right)^2\beta^2\right]\cos^2\theta~~,\\
        &\boldsymbol{h}\rightarrow \boldsymbol{h}'~~,
    \end{aligned}
\end{equation}
in all expressions obtained for states with $\bs\beta=0$.

\section{Details on the stability and causality constraints} \label{app:furtherdetails}
In this section we provide additional details on the stability and causality constraints arising in the magnetosonic channel in section \ref{sec:cM} and give an abstract procedure for how to derive them.

The Liénard-Chipart criterion provides a simpler set of necessary and sufficient condition for a high-order polynomial to be Hurwitz-stable. We can apply this criterion to the 9th degree of polynomial \eqref{Poly_P9}. This procedure schematically consists in writing the $9\times9$ Hurwitz matrix
\begin{equation}
    \begin{pmatrix}
        B_8^{(k)}&1&0&0&0&0&0&0&0 
        \\
        B_6^{(k)}&B_7^{(k)}&B_8^{(k)}&1&0&0&0&0&0
        \\
        B_4^{(k)}&B_5^{(k)}&B_6^{(k)}& B_7^{(k)}& B_8^{(k)}&1&0&0&0
        \\
        B_2^{(k)}&B_3^{(k)}&B_4^{(k)}& B_5^{(k)}& B_6^{(k)}&B_7^{(k)}&B_8^{(k)}&1&0
        \\
        B_0^{(k)}&B_1^{(k)}&B_2^{(k)}&B_3^{(k)}&B_4^{(k)}& B_5^{(k)}& B_6^{(k)}&B_7^{(k)}&B_8^{(k)}
        \\
        0&0&B_0^{(k)}&B_1^{(k)}&B_2^{(k)}&B_3^{(k)}&B_4^{(k)}&B_5^{(k)}&B_6^{(k)}
        \\
        0&0&0&0&B_0^{(k)}&B_1^{(k)}&B_2^{(k)}&B_3^{(k)}&B_4^{(k)}
        \\
        0&0&0&0&0&0&B_0^{(k)}&B_1^{(k)}&B_2^{(k)}
        \\
        0&0&0&0&0&0&0&0&B_0^{(k)}
    \end{pmatrix}
\end{equation}
and imposing the following set of conditions
\begin{equation}
\label{eq: MS_k_stability}
    \begin{split}
    &B_0^{(k)}>0~,~ B_2^{(k)}>0~,~ B^{(k)}_4>0~,~ B^{(k)}_6>0~, ~B_8^{(k)}>0,
    \\
    &\Delta_3>0~,~ \Delta_5>0~,~ \Delta_7>0~,~ \Delta_8>0~~,
    \end{split}
\end{equation}
where $\Delta_n$ labels the $n\times n $ principal minor of the Hurwitz matrix above. We provide the entries in this matrix in an ancillary Mathematica file. 

We now focus on the causality constraints on the 4th order polynomial \eqref{eq:Polycausal}. A polynomial is said to be Schur stable if all its roots lies inside the open unit disk in the complex plane and this is precisely the condition we need for causality, in addition to the requirement that $W$ is real. Schur stability can be implemented after imposing the Routh-Hurwitz criterion on the M\"{o}bius-transformed polynomial, namely
\begin{equation}
    Q^{(\infty)}_4(z)\equiv(z-1)^4 P^{(\infty)}_4\left(\frac{z+1}{z-1}\right)~~.
\end{equation}
Expressed in the form $Q^{(\infty)}_4(z)=C_4^{(\infty)} z^4+C_3^{(\infty)} z^3+C_2^{(\infty)} z^2+C_1^{(\infty)}z+C_0^{(\infty)}$ the coefficients read as the following combinations of the $B^{(\infty)}_n$:
\begin{equation}
\label{eq: BtoC}
\begin{split}
    C_4^{(\infty)}&=1 + B^{(\infty)}_1 + B^{(\infty)}_3 + B^{(\infty)}_5 + B^{(\infty)}_7,
    \\
    C_3^{(\infty)}&=4 - 4  B^{(\infty)}_1 - 2  B^{(\infty)}_3 + 2  B^{(\infty)}_7,
    \\
    C_2^{(\infty)}&=6 + 6 B^{(\infty)}_1 - 2 B^{(\infty)}_5,
    \\
    C_1^{(\infty)}&=4 - 4 B^{(\infty)}_1 + 2 B^{(\infty)}_3 - 2 B^{(\infty)}_7,
    \\
    C_0^{(\infty)}&=1 + B^{(\infty)}_1 - B^{(\infty)}_3 + B^{(\infty)}_5 - B^{(\infty)}_7.
\end{split}
\end{equation}
Assuming $C^{(\infty)}_4>0$ the Routh-Hurwitz criterion applied to $Q_4^{(\infty)}(z)$ demands
\begin{equation}
\label{eq: MS_Schur_causality}
    C^{(\infty)}_3>0,\quad C^{(\infty)}_0>0, \quad \left(C^{(\infty)}_3 C^{(\infty)}_2-C^{(\infty)}_4 C^{(\infty)}_1\right)C^{(\infty)}_1-\left(C^{(\infty)}_3\right)^2 C^{(\infty)}_0>0~~,
\end{equation}
which by means of eqs. \eqref{eq: BtoC} can be converted in relations for the coefficients $B^{(\infty)}_n$.
These ensure that $|W|^2<1$, whereas the condition $W^2>0$ requires the discriminant of $P_4^{(\infty)}(W^2)$ to be non-negative. The lengthy expression of the discriminant of the polynomial at hand is 
\begin{equation}
\label{eq: MS_real_causality}
\begin{split}
    &256 \left(B^{(\infty)}_0 B^{(\infty)}_4\right)^3-192 \left(B^{(\infty)}_0 B^{(\infty)}_4\right)^2B^{(\infty)}_3 B^{(\infty)}_1-128\left(B^{(\infty)}_0 B^{(\infty)}_4B^{(\infty)}_2\right)^2\\
    &+144\left(B^{(\infty)}_0 B^{(\infty)}_1\right)^2B^{(\infty)}_2 B^{(\infty)}_0
    \\
    &-27 \left(B^{(\infty)}_4\right)^2 \left(B^{(\infty)}_1\right)^4+144 \left(B^{(\infty)}_3 B^{(\infty)}_0\right)^2B^{(\infty)}_4 B^{(\infty)}_2-6\left(B^{(\infty)}_3 B^{(\infty)}_1\right)^2B^{(\infty)}_0 B^{(\infty)}_4
    \\
    &-80 B^{(\infty)}_0 B^{(\infty)}_4B^{(\infty)}_3 B^{(\infty)}_1\left(B^{(\infty)}_2\right)^2+18 B^{(\infty)}_4B^{(\infty)}_3 B^{(\infty)}_2\left(B^{(\infty)}_1\right)^3+16 B^{(\infty)}_0 B^{(\infty)}_4  \left(B^{(\infty)}_2 \right)^4\\
    &-4 B^{(\infty)}_0 \left(B^{(\infty)}_2\right)^3\left(B^{(\infty)}_1\right)^2-27\left[ \left(B^{(\infty)}_3\right)^2 B^{(\infty)}_0\right]^2+18  B^{(\infty)}_0 B^{(\infty)}_2 B^{(\infty)}_1\left( B^{(\infty)}_3\right)^3
    \\
    &-4\left(B^{(\infty)}_3\right)^2 \left(B^{(\infty)}_1\right)^3-4\left(B^{(\infty)}_3\right)^2\left(B^{(\infty)}_2\right)^3B^{(\infty)}_0+ \left(B^{(\infty)}_1B^{(\infty)}_3B^{(\infty)}_2\right)^2>0.
\end{split}
\end{equation}
We provide the coefficients in eq.~\eqref{eq: BtoC} in the ancillary Mathematica file.

\subsection{Conditions for a convenient choice of frame}
The stability and causality analysis, which we outlined above in full generality, is considerably simplified by restricting to the longitudinal and transverse directions $\theta=0,\pi/2$ and by picking the frame \eqref{MS_choice1}. 

\paragraph{Stability and causality for $\theta=0$.}

Along the longitudinal direction $\theta=0$ the generic magnetosonic spectral function \eqref{Poly_P9} factorizes into a $4$th and a $5$th degree polynomial $P_9(\Delta)=P_4(\Delta)P_5(\Delta).$ The $4$th-degree polynomial is the same as the Alfvén channel polynomial in \eqref{Alfvén_poly} evaluated at $\theta=0$. The conditions for stability and causality of $P_4(\Delta)$, therefore, follow from those imposed on the Alfvén channel, namely eqs. \eqref{eq:cc1},\eqref{eq: Alfv_stability} and \eqref{eq: Alfv_causality}. In turn, the $5$th-degree polynomial instead reads
\begin{equation}
\begin{split}
    P_5(\Delta)&=\Delta^5+b_4 \Delta^4+b_3 \Delta^3 + b_2 \Delta^2+b_1 \Delta+b_0~~,
\end{split}
\end{equation}
with the explicit expressions for the coefficients given by
\begin{equation}
\label{eq: P5_coeffs}
    \begin{split}
         b_4&=-\left(\frac{c T^2+2T\lambda \mu +\mu \chi}{T \varepsilon_3}+\frac{T\chi}{\varrho_4}+\frac{s T}{\chi_1}\right)~~,
         \\
         b_3&=\frac{s T}{\chi_1} \left(\frac{c T^2+2T\lambda \mu +\mu \chi}{T \varepsilon_3}+\frac{T\chi}{\varrho_4}\right)+\frac{T^2 \left(c \chi -\lambda ^2\right)}{\varepsilon_3 \varrho_4}+k^2\left(\frac{\varepsilon_2}{T \varepsilon_3}+\frac{\tau_3}{\varepsilon_3}-\frac{\tau_2}{\chi_1}+\frac{\varrho_2 \tau_4}{\varrho_4 \chi_1}+\frac{T \tau_3}{\chi_1}\frac{\varepsilon_2}{T \varepsilon_3}\right)~~,
         \\
         b_2&=-\frac{s T}{\chi_1}\frac{T^2(c\chi-\lambda^2)}{\varepsilon_3 \varrho_4}+k^2\left[-\frac{s}{\varepsilon_3}+\frac{T \lambda+\mu \chi}{T \varepsilon_3}\left(\mu+\frac{\tau_4}{\varrho_4}+\frac{T\varrho_2}{\varrho_4}\right)+\frac{\varepsilon_2}{T \varepsilon_3}\frac{(T\lambda+\mu\chi)}{\chi_1}\left(\mu + \frac{\tau_4}{\varrho_4}+\frac{T\varrho_2}{\varrho_4}\frac{T\varepsilon_3}{\varepsilon_2}\right)+\right.
         \\
         &\hspace{10mm}+\left.\frac{cT^2+2T\lambda\mu+\mu^2\chi}{T\varepsilon_3}\left(\frac{\tau_2}{\chi_1}-\frac{\varrho_2\tau_4}{\varrho_4\chi_1}\right)-\left(\frac{T\chi}{\varrho_4}+\frac{s T}{\chi_1}\right)\left(\frac{\varepsilon_2}{T\varepsilon_3}+\frac{\tau_3}{\varepsilon_3}\right)+\frac{T\chi}{\varrho_4}\left(\frac{\mu\rho_2}{\chi_1}+\frac{\tau_2}{\chi_1}-\frac{T\tau_3}{\chi_1}\frac{\varepsilon_2}{T\varepsilon_3}\right)\right]~~,
         \\
         b_1&=k^4\left(\frac{\tau_2}{T \varepsilon_3}-\frac{\varrho_2 \tau_4}{T\varrho_4 \varepsilon_3}\right)+k^2\left[\frac{T\chi}{\varrho_4}\left(\frac{s}{\varepsilon_3}+\frac{s T}{\chi_1}\left(\frac{\varepsilon_2}{T \varepsilon_3}+\frac{\tau_3}{\varepsilon_3}\right)\right)-\frac{T^2(c\chi-\lambda^2)}{\varepsilon_3\varrho_4}\left(\frac{\mu \varrho_2}{\chi_1}+\frac{\tau_2}{\chi_1}\right)+\right.
         \\
         &\hspace{80mm}+\left.\frac{s T}{\chi_1}\left(\frac{s}{\varepsilon_3}-\frac{(T\lambda+\mu \chi)}{T\varepsilon_3}\left(\mu+\frac{\tau_4}{\varrho_4}+\frac{T \varrho_2}{\varrho_4}\right)\right)\right]~~,
         \\
         b_0&=-k^2\frac{T \chi}{\varrho_4}\left[\frac{s T}{\chi_1}\frac{s }{\varepsilon_3}+k^2\left(\frac{\mu\varrho_2}{T\varepsilon_3}+\frac{\tau_2}{T\varepsilon_3}\right)\right]~~.
    \end{split}
\end{equation}
As in the most general case discussed at the beginning of this appendix, one begins by constructing the associated $5\times5$ Hurwitz matrix
\begin{equation}
    \begin{pmatrix}
        b_4 & 1 & 0 & 0 & 0\\
        b_2 & b_3 & b_4 & 1 & 0\\
        b_0 & b_1 & b_2 & b_3 & b_4\\
        0 & 0 & b_0 & b_1 & b_2\\
        0 & 0 & 0 & 0 & b_0
    \end{pmatrix}~~.
\end{equation}
For this polynomial to be stable the Liénard-Chipart criterion demands the following inequalities to hold
\begin{equation}
    \label{eq: Stability_P5}
    b_0>0,~~b_2>0,~~b_4>0,~~\Delta_3>0,~~\Delta_4>0~~,
\end{equation}
where $\Delta_{n}$ is the $n\times n $ principal minor of the Hurwitz matrix above. Given the requirements \eqref{eq:cc5} the condition $b_0>0$ leads to the sufficient conditions $\rho_2<0$ and $\tau_2<0$. In turn $b_4>0$ requires $\chi_1\lesssim(\varepsilon_3,\varrho_4)$. Additionally, we must impose the causality conditions on $P_5(\Delta)$. In the limit $k\to \infty $ one has $P^{(\infty)}_5(W)=(-i)W k^5 P^{\infty}_2(W^2)+\mathcal{O}(k^4)$ with
\begin{equation}
    P^{(\infty)}_2(W^2)=W^4+b^{(\infty)}_3 W^2+b^{(\infty)}_1~~,
\end{equation}
where $W$ is the phase velocity in the small-wavelength regime and the coefficients are related to the leading behaviour of the coefficients \eqref{eq: P5_coeffs} at large $k$. $P_2^{(\infty)}(W^2)$ is a 2nd degree polynomial in $W^2$ 
for which the causality conditions are 
\begin{equation}
    \label{eq: Causality_P5}
        \left(b^{(\infty)}_3\right)^2-4b^{(\infty)}_1>0,~~ b^{(\infty)}_3<0, ~~ b^{(\infty)}_1+b^{(\infty)}_3 +1>0 ~~,
\end{equation}
and where the first condition ensures $W^2>0$ while the other two are needed for imposing $W^2<1$.

\paragraph{Stability and causality for $\theta=\pi/2$.}
We now study the transverse directions under the same assumptions \eqref{MS_choice1}. 
The spectral function factorizes according to $P_9(\Delta)=P_6(\Delta)P_3(\Delta)$.
The explicit expressions for $P_3(\Delta)=\Delta^3+c_2 \Delta^2+c_1 \Delta +c_0$ are
\begin{equation}
\label{eq: P3_coeffs}
\begin{split}
    c_2&=-\frac{\rho}{n_2\mu}-\frac{s T}{\chi_1}~~,
    \\  
    c_1&=\frac{\rho}{n_2\mu}~\frac{s T}{\chi_1}+k^2\left(\frac{\ell_2 n_1}{n_2 \chi_1}-\frac{\ell_1}{\chi_1}\right)~~,
    \\
    c_0&=k^2\frac{\rho}{\mu n_2 \chi_1}\left(\mu \ell_2+n_1 \mu-\ell_1-\mu^2 n_2\right)~~.
\end{split}
\end{equation}
The Routh-Hurwitz criterion implies stability whenever
\begin{equation}
    \label{eq: P3_Stability}
    c_2>0,~~ c_0>0 ,~~ c_2~ c_1-c_0>0~~.
\end{equation}
Notice that the first of these conditions is automatically satisfied by the stability condition \eqref{eq:cc5} together with thermodynamics relations \eqref{eq:assthermal}. The condition $c_0>0$ implies that $\mu \ell_2+n_1 \mu-\ell_1-\mu^2 n_2>0$. For large $k$ one has $P_3(\Delta)=(-i) k^3 W \left(W^2+c_1^{\infty}\right)+\mathcal{O}(k^2)$. Consequently, causality holds for 
\begin{equation}
    \label{eq: P3_Causality}
    1>-c_1^{(\infty)}>0~.
\end{equation}
In turn, the expression for $P_6(\Delta)$ is $P_6(\Delta)=\Delta^6+d_5 \Delta^5+d_4 \Delta^4+d_3 \Delta^3 +d_2\Delta^2+d_1\Delta +d_0$. We give the specific coefficients in the ancillary Mathematica file. For stability we need the $6\times6$ Hurwitz matrix, namely 
\begin{equation}
    \begin{pmatrix}
        d_5 & 1 & 0 & 0 & 0 & 0 \\ 
        d_3 & d_4 & d_5 & 1 & 0 & 0 \\ 
        d_1 & d_2 & d_3 & d_4 & d_5 & 1 \\ 
        0 & d_0 & d_1 & d_2 & d_3 & d_4 \\ 
        0 & 0 & 0 & d_0 & d_1 & d_2 \\ 
        0 & 0 & 0 & 0 & 0 & d_0 \\ 
    \end{pmatrix}
\end{equation}
with the Liénard-Chipart conditions that read
\begin{equation}
    \label{eq: P6_Stability}
    d_0>0,~~d_2>0,~~d_4>0,~~d_5>0,~~\Delta_3>0,~~\Delta_5>0.
\end{equation}
In the $k\to \infty $ regime the polynomial reduces to $P^{(\infty)}_6(W)=(-1) k^6 P^{\infty}_3(W^2)+\mathcal{O}(k^5)$, where
\begin{equation}
    P^{\infty}_3(W^2)=W^6 + d_4^{(\infty)}W^4 +d_2^{(\infty)} W^2+ d_0^{(\infty)} ~~,
\end{equation}
is a cubic polynomial in $W^2$. For the solution to be real one considers the associated discriminant 
\begin{equation}
    \label{eq: P6_real}
    \left(d_4^{(\infty)}~ d_2^{(\infty)}\right)^2-4 \left(d_2^{(\infty)}\right)^3-4\left(d_4^{(\infty)}\right)^3 d_0^{(\infty)}-27 \left(d_0^{(\infty)}\right)^2+18 d_0^{(\infty)} d_2^{(\infty)} d_4^{(\infty)}>0.
\end{equation}
In order to enforce the phase velocity to lie inside the unit circle, one takes into account the polynomial obtained by means of the M\"{o}bius transformation
\begin{equation}
     Q^{(\infty)}_3(z)\equiv(z-1)^3 P^{(\infty)}_3\left(\frac{z+1}{z-1}\right).
\end{equation}
The coefficients of $Q^{(\infty)}_3(z)=e_3^{(\infty)}z^3+e_2^{(\infty)}z^2+e_1^{(\infty)}z+e_0^{(\infty)}$ are expressed as
\begin{equation} \label{eq:ecoeff}
    \begin{split}
        e_3^{(\infty)}&=1+d_0^{(\infty)}+d_4^{(\infty)}+d_2^{(\infty)}~~,
        \\
        e_2^{(\infty)}&=3-3d_0^{(\infty)}+d_4^{(\infty)}-d_2^{(\infty)}~~,
        \\
        e_1^{(\infty)}&=3+3d_0^{(\infty)}-d_4^{(\infty)}-d_2^{(\infty)}~~,
        \\
        e_0^{(\infty)}&=1-d_0^{(\infty)}-d_4^{(\infty)}+d_2^{(\infty)}~~.
    \end{split}
\end{equation}
One has $W^2<1$ if $Q_3^{(\infty)}$ is Hurwitz stable, which in turns implies
\begin{equation}
    \label{eq: P6_Causality}
    \frac{e_2^{(\infty)}}{e_3^{(\infty)}}>0, ~~ \frac{e_0^{(\infty)}}{e_3^{(\infty)}}>0,~~e_2^{(\infty)}e_1^{(\infty)}-e_3^{(\infty)}e_0^{(\infty)}>0.
\end{equation}
We provide the coefficients in \eqref{eq:ecoeff} in the ancillary Mathematica file.

\paragraph{Polynomials for the holographic equation of state}
For the case analysed in Sec. \ref{sec:EOS} the polynomial $P_5(\delta)$ has coefficients \eqref{eq: P5_coeffs} parametrized in terms of the rations introduced in \eqref{eq: cnvrs} and takes the following form
\begin{equation}
\label{eq: P5_coeffs_G}
    \begin{split}
         b_4&=-(G_3+G_4+G_5)~~,
         \\
         b_3&=G_4 G_5+G_3(G_4+G_5)+\kappa^2\left[\frac{G_3}{3 G_4}+G_4(Z_1+Z_2+3 G_3 Z_1 Z_2)-G_3(X_3+Z_1+Z_2-G_5 Z_3 Z_4)\right]~~,
         \\
         b_2&=G_3 G_4 G_5+\frac{\kappa^2}{3}\bigg[G_3\frac{G_5}{G_4}+G_3\left(1-3G_5(X_3+Z_1+Z_2)\right)+
         \\
         &\hspace{50mm}+G_4\Big(1+3G_5(Z_1+Z_2)-3G_3(X_3-3G_5 Z_1 Z_2-G_5 Z_3 Z_4)\Big)\bigg]~~,
         \\
         b_1&=~~\frac{\kappa^2}{3}\left[G_5 G_3+G_4(G_5+G_3-3G_5 G_3 X_3)-\kappa^2\left(\frac{1}{3}-G_4(X_3+Z_11+Z_2-G_5Z_3 Z_4)\right)\right],
         \\
         b_0&=-\frac{\kappa^2}{3}G_5\left[G_3 G_4 -\kappa^2\left(\frac{1}{3}-G_4(Z_1+X_3+Z_2)\right)\right]~~.
    \end{split}
\end{equation}
In turn the polynomial $P_6(\delta)$, defined below eq. \eqref{eq: P3_Causality}, has coefficients
\begin{equation}
    \label{eq: P6_coeffs_G}
    \begin{split}
        d_5&=-(G_2+G_4+G_5),
        \\
        d_4&=G_4 G_5+G_2(G_4+G_5)+\kappa^2\left[\frac{1}{3}\Big(3G_5 X_1-G_4(F_1+F_3)\Big)+\frac{1}{3}\frac{G_2}{G_4}(1-F_1G_4)(1-F_3G_4)+\right.
        \\
        &\hspace{10mm}+\left.\frac{1}{4}\frac{G_2}{G_4}\Big(X_3+4X_2-4G_4(F_2+Y_1)(F_4+Y_2)\Big)\right],
        \\
        d_3&=-G_2 G_4 G_5+\kappa^2G_2\left[ G_4(X_I+X_2)+\frac{1}{4}\frac{G_5}{G_4} X_3-\frac{G_4}{3G_2}\Big(1+G_5(F_1+F_3-3X_1)\Big)+\frac{1}{4} G_4(X_3+4X_2)+\right.
        \\
        &\hspace{10mm}\left.-\frac{1}{3G_4}\Big(G_4+G_5-G_4 G_5\big(F_1+F_3+3(X_1+X_2)\big)\Big)-\frac{G_4}{3}G_5\Big(F_1F_3+3(F_2+Y_1)(F_4+Y_2)\Big)\right],
        \\
        d_2&=\frac{\kappa^2}{3}\left[G_4 G_5+G_2\left(G_4+G_5-\frac{3}{4}G_4G_5(X_3+4(X_1+X_2))\right)\right]+
        \\
        &\hspace{15mm}+\frac{\kappa^4}{3}\left[-\frac{1}{3}-\frac{G_4}{4}(4F_2F_4G_5-4X_2-X_3)-\frac{G_5}{4G_4}G_2X_1\Big(4-4G_4(F_1+F_3+3X_2)-3G_4X_2\Big)+\right.
        \\
        &\hspace{10mm}\left.+\frac{G_4}{3}\Big(F_1+F_3-3G_5X_1(F_1+F_3+F_1F_3G_2)\Big)+G_2G_4G_5\Big(F_1Y_2(F_2+Y_1)+F_3Y_1(F_4+Y_2)\Big)\right],
        \\
        d_1&=-\frac{\kappa^2}{3}G_2G_4G_5+\\
        &\hspace{10mm}+\frac{\kappa^4}{3}\left[\frac{G_4}{3}\Big(1-(F_1+F_3)G_4+3G_2X_1\Big)-\frac{1}{4}G_4G_5\Big(X_3+4X_2+4G_2(F_4 Y_1+Y_2(F_2+Y1))\Big)\right],
        \\
        d_0&=-\frac{\kappa^4}{3}G_2G_4G_5X_1+\frac{\kappa^6}{9}G_4(X_1-G_2Y_1Y_2)\left[1-G_4(F_1+F_3+3X_2)+\frac{3}{4}G_4X_3\right]~~.
    \end{split}
\end{equation}
This completes the details of the various polynomials.
%



\newpage
\providecommand{\href}[2]{#2}\begingroup\raggedright\endgroup


\end{document}